%% file: main.tex
\documentclass[journal]{IEEEtran}

\usepackage{soul,framed} 
\usepackage{graphicx}
\usepackage[table]{xcolor}
\colorlet{shadecolor}{yellow}

\graphicspath{{../pdf/}{../jpeg/}}
\DeclareGraphicsExtensions{.pdf,.jpeg,.png}

\usepackage[cmex10]{amsmath}
\DeclareMathSizes{5.5}{5.5}{5}{5}
\DeclareMathSizes{5.6}{5.6}{5}{5}
\usepackage{array}
\usepackage{mdwmath}
\usepackage{tikz}
\usepackage{eqparbox}
\usepackage{url}
\usepackage{algorithm}
\usepackage{algorithmic}
\usepackage{cite}
\usepackage[nolist]{acronym}
\usepackage{booktabs}
\usepackage{multirow}
\usepackage{multicol}
\usepackage{xltabular}
\usepackage{amsfonts}
\usepackage{pdflscape}
\usepackage{float}
\usepackage{bbm}
\usepackage{balance}
\usepackage{wrapfig} 
\usepackage{makecell}
\usepackage{amssymb}
\usepackage{pifont}

\input{acro}

\begin{document}

    \title{Multimodal Target Speaker Extraction: Towards Unified Speaker Cues Across Modalities}
    \author{Xinyuan Qian, Yanghao Zhou, Ziyang Jiang, Yu Chen, Xinjia Zhu, Xueyan Chen, \\Qiquan Zhang,~\IEEEmembership{Member~IEEE}, Zexu Pan, Jiaying Wang, Xianghu Yue, Jiadong Wang, \\Bj{\"o}rn Schuller,~\IEEEmembership{Fellow~IEEE} and Haizhou Li,~\IEEEmembership{Fellow~IEEE}
     \thanks{Xinyuan Qian, Ziyang Jiang, Xueyan Chen are with the School of Computer and Communication Engineering, University of Science and Technology Beijing, 100083, China (e-mail: qianxy@ustb.edu.cn; \{M202410665, M202510650\}@xs.ustb.edu.cn).}
     \thanks{Yanghao Zhou is with Department of Computer Science and Technology, Beijing Institute of Technology, China. e-mail: zhouyh77@bit.edu.cn}
     \thanks{Yu Chen is with the Chinese University of Hong Kong, Shenzhen, China (email: yuchen2@link.cuhk.edu.cn)}
     \thanks{Qiquan Zhang and Zexu Pan are with Alibaba Token Foundry, Alibaba Group (e-mail: zhang.qiquan@outlook.com; zexu.pan@alibaba-inc.com).}
      \thanks{Jiaying Wang is with the Zhejiang Institute of Quality Sciences (Technology Innovation Center of the State Administration for Market Regulation), Hangzhou, 310018, China (e-mail: jywangsusan@icloud.com).}
     \thanks{Xianghu Yue is with the School of Computer Software, Tianjin University, Tianjin, 300350, China (e-mail: yuexianghu@tju.edu.cn).}
     \thanks{Jiadong Wang and Bj{\"o}rn Schuller are with the University Hospital rechts der Isar, Technical University of Munich, Munich, Germany (e-mail: jiadong.wang@tum.de; schuller@tum.de).}
     \thanks{Haizhou Li is with the School of Artificial Intelligence, the Chinese University of Hong Kong, Shenzhen, 518172, China (e-mail: haizhouli@cuhk.edu.cn).}

}
\bstctlcite{IEEEexample:BSTcontrol}
    
\maketitle

\begin{abstract}
\ac{TSE} is  pivotal in speech communication and human–computer interaction, enabling the isolation of a specific speaker’s voice from complex acoustic environments, \emph{i.e.}, the cocktail party scenario. Although traditional TSE systems conditioned on enrollment speech have progressed substantially, enrollment speech as a cue has inherent limitations. Its
reliability degrades when the target and interfering speakers have similar voice characteristics, when intra-speaker variability (e.g. changes in emotion or speaking style) creates a mismatch between the enrollment and target speech, or when the enrollment itself is contaminated by noise or competing speakers.
This review surveys deep-learning-based TSE from the perspective of auxiliary target cues drawn from multiple modalities. We organize existing methods according to five types of information used to isolate the target speaker: audio enrollment, visual, spatial, textual/semantic, and neural cues.

We also trace the evolution from discriminative estimators to variational, diffusion, flow, codec, and foundation-model-based systems and summarize representative datasets and evaluation metrics. We review the benefits and limitations of different cues and discuss challenges involving synchronization, missing or unreliable observations, data scarcity, privacy, computational cost, and real-time operation.
Finally, we summarize future directions concerning adaptive cue fusion, instruction-driven extraction, realistic evaluation, and trustworthy deployment. By jointly reviewing cue design, model architecture, training objectives, datasets, and evaluation metrics, this article provides an overview of the current landscape and open problems in multimodal TSE.

\end{abstract}

\input{sections/01_introduction}

\input{sections/02_problem_formulation}

\input{sections/03_auxiliary_target_cues}

\input{sections/04_network_evolution}

\input{sections/05_datasets}

\input{sections/06_evaluation_metrics}

\input{sections/07_challenges_open_problems}

\input{sections/08_future_directions}

\input{sections/09_conclusion}

\ifCLASSOPTIONcaptionsoff
  \newpage
\fi

\bibliographystyle{IEEEtran}
\bibliography{IEEEabrv,Bibliography}

\appendix

\subsection{Taxonomy of Multimodal TSE}

This appendix summarizes the methods in Sec.~\ref{sec:type_cues} and organizes them according to the auxiliary prior information used to identify the target speaker. As shown in Table~\ref{tab:sota}, existing methods are grouped into five categories: audio enrollment and reference-conditioned systems, visual and audio-visual systems, spatial and location-conditioned systems, textual or semantic systems, and neural and attention-conditioned systems. For each method, we report the auxiliary target cue, the evidence used for target selection, and the system scope. This cue-centric taxonomy reflects the progression of TSE from speaker-identity-based conditioning toward more flexible target extraction paradigms that exploit articulatory, spatial, semantic, contextual, and listener-dependent information, either individually or jointly. Methods marked with $\dagger$ denote adjacent or unified frameworks whose primary objective is not necessarily TSE, but whose conditioning or target-selection mechanisms are transferable to multimodal TSE.

\input{table_method}

\vfill

\newpage

\end{document}

%% file: acro.tex
\newacro{TSE}[TSE]{Target Speaker Extraction}
\newacro{ASR}[ASR]{Automatic Speech Recognition}
\newacro{SE}[SE]{Speech Enhancement}
\newacro{SS}[SS]{Speech Separation}
\newacro{SNR}[SNR]{Signal-to-Noise Ratio}
\newacro{SDR}[SDR]{Signal-to-Distortion Ratio}
\newacro{SDRi}[SDRi]{Signal-to-Distortion Ratio improvement}
\newacro{LN}[LN]{Layer Normalization}
\newacro{LLM}[LLM]{Large Language Model}
\newacro{SI-SDR}[SI-SDR]{Scale-invariant Signal-to-Distortion Ratio}
\newacro{SI-SDRi}[SI-SDRi]{Scale-invariant Signal-to-Distortion Ratio improvement}
\newacro{PESQ}[PESQ]{Perceptual Evaluation of Speech Quality}
\newacro{STOI}[STOI]{Short-Time Objective Intelligibility}
\newacro{RNN}[RNN]{Recurrent Neural Network}
\newacro{DPRNN}[DPRNN]{Dual-Path Recurrent Neural Network}
\newacro{ASR}[ASR]{automatic speech recognition}
\newacro{MP-SENet}[MP-SENet]{}
\newacro{PIT}[PIT]{Permutation Invariant Training}
\newacro{OpenUnmix}[OpenUnmix]{Open-Source Music Separation}
\newacro{TF-GridNet}[TF-GridNet]{TensorFlow GridNet}
\newacro{Conv-TasNet}[Conv-TasNet]{Convolutional Time-domain Audio Separation Network}
\newacro{MP-SENet}[MP-SENet]{Multi-Path Squeeze-and-Excitation Network}
\newacro{TF}[TF]{Time-Frequency}
\newacro{DOA}[DoA]{Direction-of-Arrival}
\newacro{RIR}[RIR]{Room Impulse Responses}
\newacro{STFT}[STFT]{Short-Time Fourier Transform}
\newacro{IPD}[IPD]{Inter-channel Phase Difference}
\newacro{CLAP}[CLAP]{Contrastive Language-Audio Pretraining}
\newacro{FiLM}[FiLM]{Feature-wise Linear Modulation}
\newacro{SELD}[SELD]{Sound Event Localization and Detection}
\newacro{ILD}[ILD]{Interaural Level Differences}
\newacro{MLP}[MLP]{Multi-Layer Perceptron}
\newacro{TCN}[TCN]{Temporal Convolutional Network}

%% file: sections/01_introduction.tex
\section{Introduction}

Speech perception in real-world environments is fundamentally challenged by the coexistence of multiple speakers, background noise, and acoustic reverberation, a phenomenon commonly known as the cocktail party problem~\cite{cherry1953some}. Speech enhancement (SE) aims to reconstruct the clean speech in the presence of background noise~\cite{overview2018,DeepMMSE,Mambaspeech,10955264,tfaj}, while speech separation (SS) attempts to recover all individual sources from a mixture~\cite{hershey2016deep,yu2017permutation,convtasnet}. However, many practical applications do not require all speakers. Instead, the goal is often to selectively attend the speaker of interest while ignoring others. Target speaker extraction (TSE) addresses this problem by introducing auxiliary information to specify the desired speaker. This capability has become increasingly important in personalized communication, hearing assistance, automatic speech recognition (ASR), teleconferencing, and human-machine interaction.

Early TSE systems primarily rely on an enrollment utterance as the target cue. Representative approaches, including SpeakerBeam~\cite{speakerbeam}, VoiceFilter~\cite{wang2019voicefilter}, and SpEx~\cite{xu2020spex}, learn speaker-aware representations from reference speech and condition an extraction network on the target identity. Such audio enrollment provides direct speaker information and has driven substantial progress in TSE. However, enrollment speech as a target cue has inherent limitations. Its discriminative power can diminish when the target and interfering speakers exhibit similar voice characteristics, increasing the risk of speaker confusion. In addition, intra-speaker variability, such as changes in emotion or speaking style, can create a mismatch between the enrollment utterance and the target speech in the mixture, making the learned speaker representation less reliable. The enrollment itself may also be unavailable in spontaneous interactions or contaminated by noise or competing speakers. These limitations motivate a fundamental question: \textit{can target speakers be identified and extracted using information beyond reference speech?}

Humans rarely rely on acoustic signals alone when attending to a speaker in complex environments. Instead, selective listening naturally integrates complementary sensory and cognitive cues, such as co-speech visual motions, spatial location, semantic context, and attentional information, to focus on the desired speaker~\cite{golumbic2013visual,crosse2016eye}. Inspired by this capability, recent TSE research has gradually evolved from conventional speaker enrollment toward multimodal target specification. 

For instance, visual cues, such as facial appearance, synchronized lip movements, and gestures, provide complementary identity and articulatory information that remains informative under severe acoustic interference~\cite{ochiai2019multimodal,pan2021muse,pan2022seg,li2024momuse}. 
Spatial cues exploit the direction, distance, or multichannel localization information of the target speaker to distinguish competing sources occupying different positions~\cite{gu2019neural,gu2020multi,ge2022spex}. 
More recently, language prompt cues have enabled target specification through semantic descriptions, speaker attributes, or inter-speaker relationships, offering a more flexible and natural interaction paradigm~\cite{hao2025typing,seki2025language,dai2025inter}. 
Neural cues, such as electroencephalography (EEG), infer the listener's auditory attention directly from brain responses, providing a promising direction for hearing-assistive and brain--computer interface applications~\cite{pan2023neuroheed,neurospex}. These auxiliary cues characterize the target speaker from complementary perspectives. Together, they substantially broaden the way to specify the target, extending TSE beyond conventional enrollment-based systems.

\begin{figure}[!t]
  \centering
    \includegraphics[width=\linewidth]{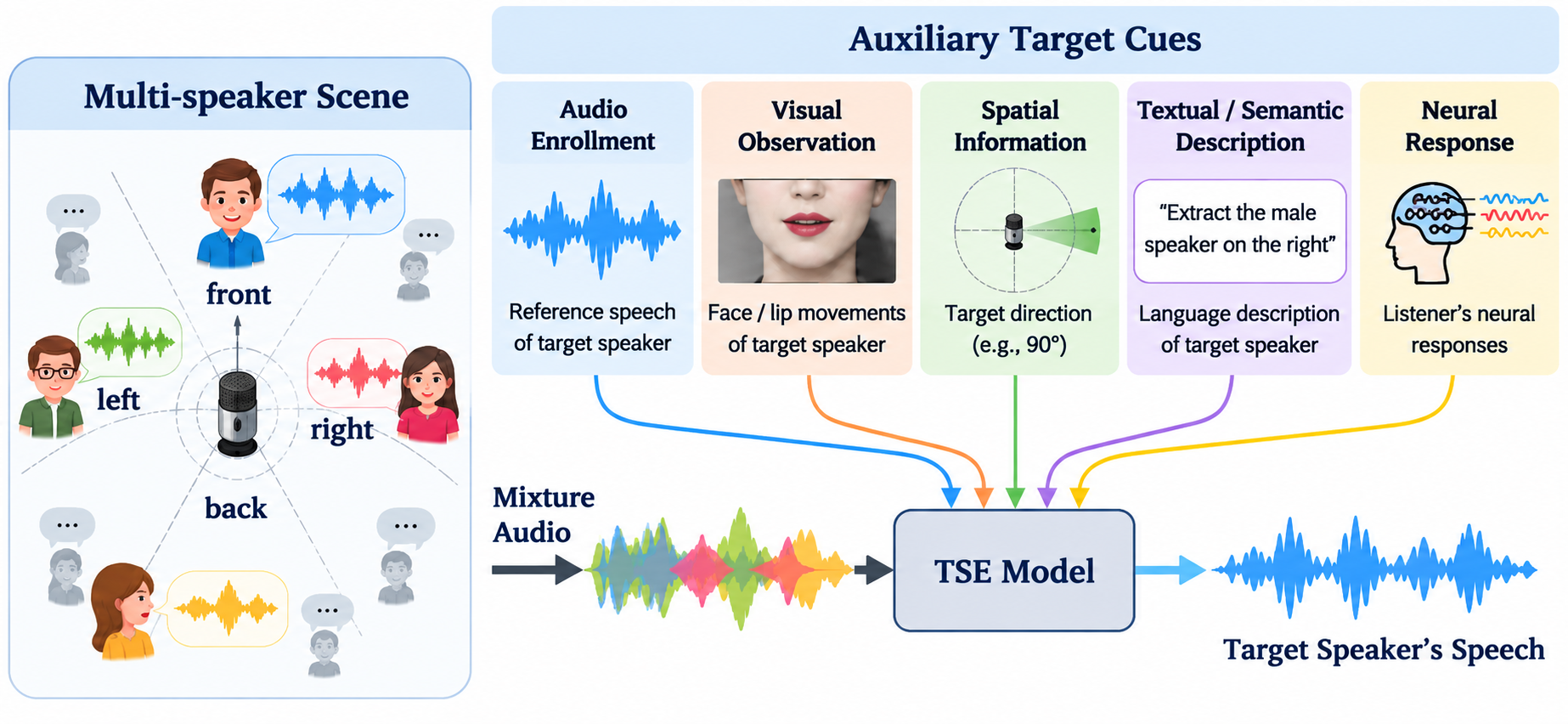}
  \caption{Illustration of TSE guided by different auxiliary target cues i.e., audio enrollment, visual observations, spatial information, textual or semantic descriptions, neural responses, or combinations of multiple cues.}
  \label{Cover}
\end{figure}

In this review, we organize multimodal TSE methods from the perspective of  \textit{how the target speaker is specified}. Here, ``multimodal'' refers to the diversity of auxiliary cues explored in TSE, rather than requiring every system to combine all modalities. 
Figure~\ref{Cover} summarizes five cue categories considered in this review: audio cues (e.g., enrollment speech), visual cues (e.g., face and lip movements), spatial cues (e.g., speaker location), textual or semantic cues (e.g., language queries or instructions), and neural cues (e.g., electroencephalography signal). These cues differ not only in the target information type, but also in their sensing requirements, synchronization constraints, and robustness under real-world conditions.

Beyond the diversity of target speaker's cues, TSE models have also undergone substantial architectural evolution. Early solutions were dominated by discriminative approaches, which mainly include time-frequency mask estimation~\cite{10389752} and time-domain waveform extraction based on an encoder-separator-decoder architecture~\cite{convtasnet}. More recently, probabilistic generative formulations have been introduced, including diffusion models, variational inference, and flow-matching, enabling more flexible modeling of speech~\cite{wang2022directioncvae,wang2024dualcvae,difftse,ddtse,flowtse}. Meanwhile, advances in neural audio codecs, discrete speech representations, and large-scale pretrained models have opened new directions toward unified and foundation-model-based speech processing systems~\cite{skim,tselm,lauratse,speechx,sslmhfa,nemo,uniaudio,wang2025metis,llaseg1,anyenhance}. Together, these developments indicate a transition of TSE from task-specific discriminative models toward multimodal, generative, and pretrained architectures.

\begin{figure*}[!t]
  \centering
  \includegraphics[width=0.99\linewidth]{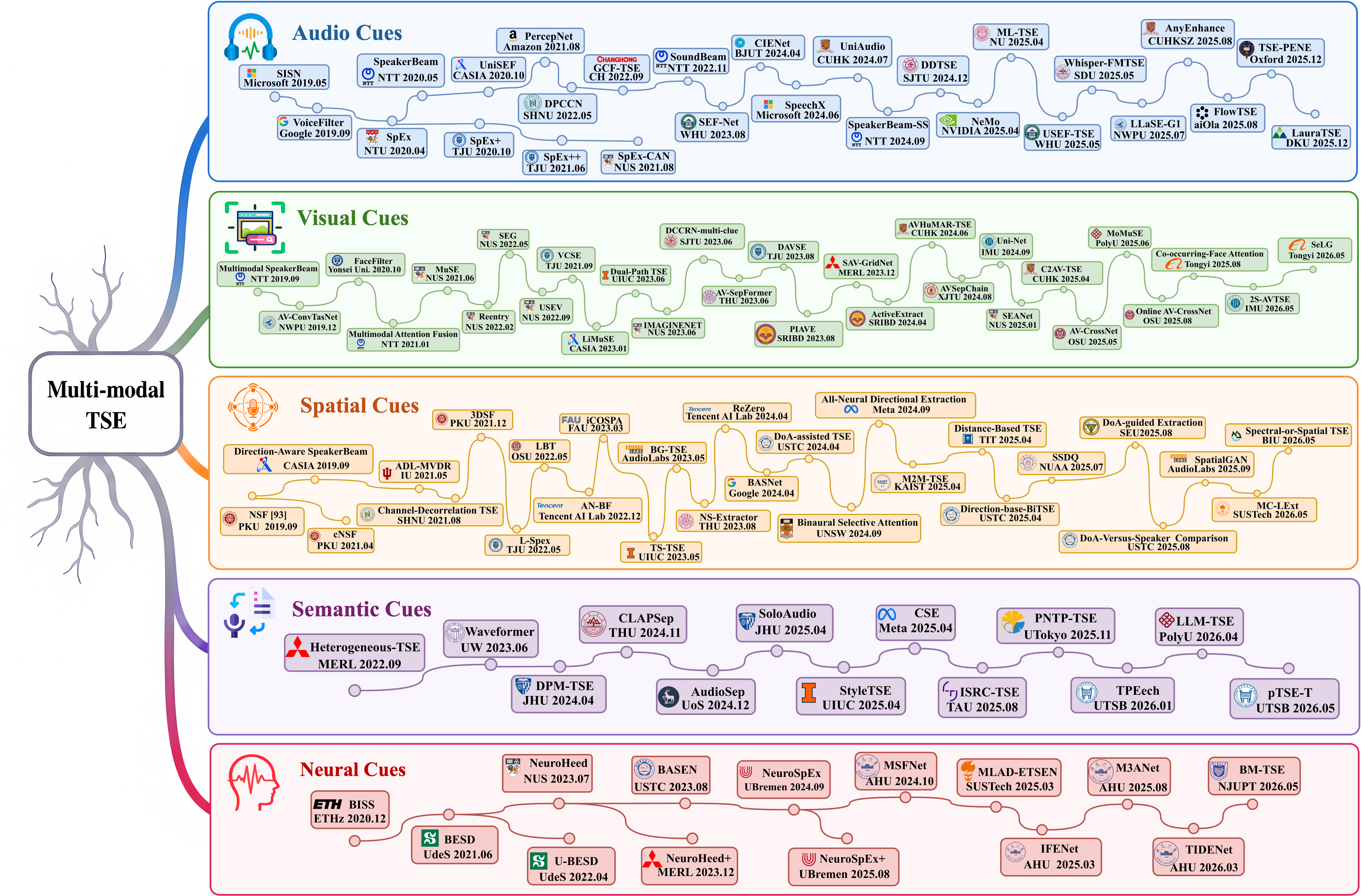}
  \caption{Taxonomy of TSE methods according to auxiliary target cues e.g., audio, visual, spatial, semantic and neural information.
  }
  \label{fig:evolved_taxonomy}
\end{figure*}

This review provides a unified view of TSE along two dimensions: \textit{what information identifies the target speaker} and \textit{how the extraction process is modeled}. 
We focus on auxiliary cues for specifying, describing, localizing, or inferring the target speaker, while incorporating related sound restoration and generation studies that provide relevant conditioning or modeling insights. Unlike general source separation reviews, we trace the evolution of target-conditioned extraction across sensing modalities (as summarized in Figure~\ref{fig:evolved_taxonomy}) and model architectures. The main contributions  are listed as follows:
\begin{itemize}
\item {We present a unified taxonomy of TSE methods organized around five target cues, i.e., audio, visual, spatial, textual/semantic, and neural, revealing how different cues are represented and integrated.}

\item {We review the evolution of TSE, charting the shift from discriminative formulations to generative paradigms, including diffusion, flow-matching, and auto-regressive approaches.}

\item We summarize datasets and evaluation protocols spanning simulated mixtures, controlled recordings, and multimodal benchmarks, and analyze their suitability for practical scenarios.

\item We highlight open challenges and future directions, including missing cues, domain generalization, interactive selection, efficiency, evaluation, and privacy.
\end{itemize}

The remainder of this paper is organized as follows. Sec.~\ref{sec:problem_formulation} formulates the multimodal TSE. Sec.~\ref{sec:type_cues} reviews the methods categorized by target cues. Sec.~\ref{sec:evolution} discusses architecture evolution. Sec.~\ref{sec:dataset} and~\ref{sec:evaluation_metrics} summarize datasets and evaluation protocols, followed by discussions on current challenges and future directions in Sec.~\ref{sec:challenge} and~\ref{sec:future_direction}.

%% file: sections/02_problem_formulation.tex
\section{Problem Formulation}
\label{sec:problem_formulation}

In real-world acoustic environments, the speech signal captured by microphones usually contains multiple concurrent sources, including the target speaker, interfering speakers, and background noise. TSE aims to recover the speech signal of a desired speaker from such a mixture by exploiting additional information that identifies the target speaker.

Formally, the observed mixture signal at the $m$-th microphone can be formulated as
\begin{equation}
    \mathbf{x}^{m}(n)
    =
    \mathbf{s}^{m}_{t}(n)
    +
    \sum_{i\neq t}\mathbf{s}^{m}_{i}(n)
    +
    \mathbf{d}^{m}(n),
\end{equation}
where $\mathbf{s}^{m}_{t}(n)$ denotes the target speaker's speech signal, $\mathbf{s}^{m}_{i}(n)$ represents the interfering speech signal from the $i$-th speaker, $\mathbf{d}^{m}(n)$ denotes background noise, $n$ is the discrete time index, $t$ is the target speaker index, and $m$ is the microphone index. We use $\mathbf{x}=\{\mathbf{x}^{m}\}_{m=1}^{M}$ to denote the available microphone recordings; the single-channel case corresponds to $M=1$.

Given an auxiliary target cue $\mathbf{Q}_{t}$, a TSE system estimates the target speech signal from the mixture as
\begin{equation}
    \hat{\mathbf{s}}_{t}
    =
    f_{\theta}^{\mathrm{TSE}}
    (\mathbf{x},\mathbf{Q}_{t}),
    \label{eq:general_tse}
\end{equation}
where $\hat{\mathbf{s}}_{t}$ is the estimated target speech, with its output-channel structure determined by the system, and
$f_{\theta}^{\mathrm{TSE}}(\cdot)$ represents a TSE model parameterized by $\theta$.
The auxiliary cue $\mathbf{Q}_{t}$ provides information specifies the target source, distinguishing it from competing speakers and background interference.

Conventional TSE approaches mainly rely on pre-enrolled speech as the target cue, while recent studies explore additional visual, spatial, textual/semantic, and neural cues. We define the target cue set as
\begin{equation}
\mathbf{Q}_{t}
=
\{
\mathbf{Q}^{a}_{t},
\mathbf{Q}^{v}_{t},
\mathbf{Q}^{sp}_{t},
\mathbf{Q}^{sm}_{t},
\mathbf{Q}^{ne}_{t}
\},
\end{equation}

where each element denotes a target cue category, which are described below.

\begin{itemize}
\item \textbf{Audio cues} $\mathbf{Q}^{a}_{t}$ are derived from target speaker enrollment utterances. Speaker-discriminative representations i.e.,  speaker embeddings are extracted to guide TSE.

\item \textbf{Visual cues} $\mathbf{Q}^{v}_{t}$ include static and dynamic observations of the target speaker, such as facial appearance or expression, lip motion, head pose, and gestures, providing identity and articulation information.

\item \textbf{Spatial cues} $\mathbf{Q}^{sp}_{t}$ offer target-related spatial information, such as direction of arrival (DoA), distance, or region, using spatial features including inter-channel phase difference (IPD), inter-channel time difference (ITD), and beamforming outputs to guide extraction.

\item \textbf{Textual or semantic cues} $\mathbf{Q}^{sm}_{t}$ offer target-related linguistic information, such as speaker attributes, transcripts, and dialogue context, using text encoders or large language models (LLMs) to guide extraction~\cite{seki2025language,dai2025inter,hao2025typing}.

\item \textbf{Neural cues} $\mathbf{Q}^{ne}_{t}$ are derived from neural signals, such as electroencephalography (EEG), electrocorticography (ECoG), and magnetoencephalography (MEG), and reflect the listener's auditory attentional state.
\end{itemize}

For  discriminative TSE, the training objective can be generally expressed as:
\begin{equation}
    \mathcal{L}
    =
    \mathcal{L}_{\mathrm{rec}}
    +
    \alpha \mathcal{L}_{\mathrm{aux}},
    \label{eq:generic_tse_loss}
\end{equation}

where $\mathcal{L}_{\mathrm{rec}}$ denotes reconstruction objectives between $\hat{\mathbf{s}}_{t}$ and $\mathbf{s}_{t}$ (\emph{e.g.}, SI-SDR), and $\mathcal{L}{\mathrm{aux}}$ captures cue-specific objectives, such as speaker discrimination, cross-modal alignment, spatial guidance, semantic consistency, or neural conditioning. This decomposition provides a unified perspective rather than a common loss formulation.

%% file: sections/03_auxiliary_target_cues.tex
\section{TSE with Different Auxiliary Target Cues}
\label{sec:type_cues}

This section reviews how auxiliary target cues are represented and integrated into TSE systems. Existing methods are categorized into five cue types: audio, visual, spatial, textual/semantic, and neural (as listed in Table~\ref{tab:sota}), with multimodal systems discussed when multiple cues are combined. We analyze cue-specific architectures, conditioning strategies, fusion mechanisms and limitations.

\input{sections/03_1_audio_cues}

\input{sections/03_2_visual_cues}

\input{sections/03_3_spatial_cues}

\input{sections/03_4_textual_cues}

\input{sections/03_5_neural_cues}
\input{sections/03_6_crossmodal_comparison}


%% file: sections/03_1_audio_cues.tex
\subsection{Audio Cues}
Audio-conditioned TSE typically specifies the target speaker through an enrollment utterance. The enrollment utterance must first be represented in a form suitable for extraction and then used to guide the processing of the mixture. Accordingly, this subsection reviews cue encoding, conditioning mechanisms, and practical limitations. Cue encoding distinguishes explicit speaker embedding from speaker embedding-free features, whereas conditioning concerns how either type of cue interacts with the mixture. 

\subsubsection{Cue Types and Encoding}

\emph{Speaker-embedding-based encoding.} These methods encode enrollment speech into an explicit, fixed-dimensional speaker embedding, with or without speaker-identity labels. In identity-supervised systems, labels are used either to pretrain a fixed speaker encoder or to provide auxiliary classification, verification, or metric-learning objectives during joint extraction training. Typical pretrained embeddings include d-vectors, GE2E embeddings, and ECAPA-TDNN embeddings~\cite{dvector,wan2018generalized,desplanques2020ecapa}. VoiceFilter uses a separately pretrained verification encoder~\cite{wang2019voicefilter}, whereas TD-SpeakerBeam investigates auxiliary speaker-identification supervision~\cite{delcroix2020improving}, and SpEx jointly trains a mean-pooled BLSTM speaker encoder with speaker-classification and reconstruction losses~\cite{spex}. Ji \emph{et al.} combine verification pretraining with joint extraction and verification training~\cite{ji2020speakeraware}. Speaker-centroid estimation uses speaker-indexed centroid targets and an angular prototypical objective~\cite{heo2024centroid}.

Without identity labels, an explicit speaker embedding can instead be learned through extraction or self-supervised objectives. Early SpeakerBeam jointly learns a pooled speaker vector through extraction loss~\cite{delcroix2018speakerbeam,speakerbeam}; its optional deep-clustering objective supervises within-mixture source assignments rather than global speaker identities~\cite{speakerbeam,hershey2016deep}. Peng \emph{et al.} pool pretrained self-supervised WavLM features into an enrollment embedding~\cite{sslmhfa}. SDR-TSE design a self-supervised mechanism of speaker-content disentanglement to obtain a speaker embedding without identity labels~\cite{mu2024self}. These approaches remain embedding-based because they explicitly summarize the reference into a speaker vector. Likewise, the speech-discretization system of Yu \emph{et al.} retains a speaker encoder despite predicting acoustic tokens~\cite{skim}.

\emph{Speaker-embedding-free encoding.} These methods retain the reference as frame-level features or discrete tokens, or the waveform itself, rather than producing a separate global speaker embedding. Frame-level features include the contextual time--frequency features in CIENet and its coarse-to-fine extension~\cite{cienet,yang2024coarsetofine}, DCF-Net~\cite{xue2025dcf}, the convolutional features in USEF-TSE~\cite{usef}, and the Conformer features in LauraTSE, and the sixth-layer WavLM reference features in LLaSE-G1~\cite{lauratse,llaseg1}. Discrete representations include WavLM tokens in TSELM~\cite{tselm} and audio-codec prompts in SpeechX, UniAudio, and AnyEnhance~\cite{speechx,uniaudio,anyenhance}. LExt takes the third approach which retains the raw enrollment waveform without a dedicated enrollment encoder~\cite{shen2025lext}. 

\subsubsection{Conditioning Mechanisms}

Once encoded, enrollment features guide the extraction of the target speaker. Explicit speaker embeddings can be injected through concatenation, multiplication, gating, or feature-wise modulation. SpeakerBeam explores input-bias adaptation, factorized layers, and scaled activations~\cite{speakerbeam}, while GCF-TSE~\cite{gcntse} uses gated fusion to control how speaker embedding modifies mixture features. SpEx~\cite{spex} employs multi-scale architectures to adapt cue representations across resolutions. SpEx+~\cite{spex+} reduces feature-space mismatch by applying weight-shared waveform encoders separately to the mixture and enrollment before deriving the speaker embedding and extraction. SpEx-CAN~\cite{spexcan} uses cross-attention between speaker embedding and mixture representations. Hierarchical and multi-level methods introduce local and global cues at different processing levels, enabling scale-aware conditioning~\cite{he2024hierarchical,mltse}. Multi-target speaker extraction extends conditioning from one target enrollment to a requested set represented by multiple embeddings~\cite{serre2025mtse}.

Speaker-embedding-free encoding enable finer-grained mixture-enrollment conditioning rather than broadcasting one vector across all mixture frames. CIENet uses attention-based enrollment--mixture interaction to construct mixture-aligned contextual guidance~\cite{cienet}, and its coarse-to-fine extension repeats contextual interaction and extraction at successive resolutions~\cite{yang2024coarsetofine}. DCF-Net models the interactions between enrollment and mixture representations across time and frequency axes~\cite{xue2025dcf}. USEF-TSE combines enorllment--mixture cross-attention with feature-wise modulation in its time-domain implementation or time--frequency implementation~\cite{usef}, while TSELM uses cross-attention to incorporate enrollment information into target-token prediction~\cite{tselm}. 

Prompt-based conditioning, as one of Speaker-embedding-free conditioning, places the enrollment and mixture together through temporal, channel-wise composition in terms of raw wavforms or features encoded by the same encoder. LExt prepends the enrollment waveform to the mixture and processes both with a shared separation network~\cite{shen2025lext}. In generative systems, LauraTSE uses continuous mixture and enrollment sequences to condition autoregressive target codec-token prediction with a decoder-only language model~\cite{lauratse}; SpeechX, UniAudio, LLaSE-G1, and AnyEnhance similarly condition target generation on discrete or continuous audio prompts~\cite{speechx,uniaudio,llaseg1,anyenhance}. 

\subsubsection{Practical Limitations}

The usefulness of an enrollment cue depends on its distinctiveness, its consistency with the target speech in the mixture, and the quality of the reference recording. We examine methods addressing speaker confusion, intra-speaker mismatch, and enrollment contamination, together with the limitations that remain~\cite{zhang2024ortse,tsepene}.

\emph{Speaker similarity and confusion.} Confusion arises not only from similar voices but also from insufficiently discriminative reference representations. Zhao \emph{et al.} use metric learning to improve embedding discrimination and then detect confusion by comparing the extracted speech with enrollments of both the target and interferer~\cite{zhao2022target}. X-SepFormer emphasizes confusing chunks during training and reports a 14.8\% relative reduction in confusion errors on WSJ0-2mix~\cite{xsepformer}, while curriculum learning progressively introduces more similar speaker pairs and improves iSDR from 12.57 to 13.44 dB on Libri2talker~\cite{liu2024curriculum}. For errors specific to target-speaker conditioning, hybrid continuity losses~\cite{pan2022hybridcontinuity} penalize discontinuities in the extracted signal. Nevertheless, these gains depend on particular difficulty measures, mixture assumptions, and simulated benchmarks; they reduce observed confusion without guaranteeing reliable target discrimination for unseen speakers with highly similar voices.

\emph{Intra-speaker variability and mismatch.} Emotion mismatch reduces average SDRi from 14.1 to 12.4 dB on RAVDESS2Mix~\cite{vsvec2022analysis}, and normal--whispered mismatch also degrades extraction~\cite{borsdorf2024wtimit2mix}. Sato \emph{et al.} train on the most difficult enrollment among several candidates and add auxiliary speaker-identification supervision, reducing severe failures~\cite{sato2022strategies}. However, the worst case is defined only over the sampled candidates rather than unseen emotional or stylistic variations. SDR-TSE suppresses semantic and paralinguistic factors through self-supervised disentanglement~\cite{mu2024self}, whereas EvoTSE adapts the reference using reliable historical estimates~\cite{liu2026evotse}. However, disentanglement has not been validated across the full range of emotional, stylistic, vocal, and channel variations. Existing methods therefore improve robustness to selected mismatches rather than learning a universally invariant speaker cue.

\emph{Contaminated enrollment.} Noise, reverberation, or competing speech can obscure the target characteristics or introduce a wrong speaker into the enrollment cue. Enrollment augmentation with noise, reverberation, and estimated speech improves robustness to such distortions~\cite{li2024enrollmentaugmentation}. OR-TSE derives references from non-overlapping mixture regions and uses contrastive learning to suppress interfering-speaker information~\cite{zhang2024ortse}, while TSE-PENE compares a noisy positive enrollment containing the target and interfering speakers with a negative enrollment in which the target is absent, thereby isolating target-specific characteristics~\cite{tsepene}. However, augmentation covers only anticipated distortions, OR-TSE depends on diarization and sufficiently clean non-overlapping regions, and TSE-PENE degrades when the positive enrollment is severely corrupted. These methods therefore reduce sensitivity to moderate contamination but cannot recover reliable target information when noise or competing speech dominates the enrollment.

%% file: sections/03_2_visual_cues.tex
\subsection{Visual Cues}
Visual conditioning specifies the target through  speech-synchronous motions. This subsection distinguishes these forms of visual evidence, reviews their encoding and fusion mechanisms, and discusses robustness to impaired, missing, or asynchronous observations.

\subsubsection{Cue Types and Encoding}
Visual cues can identify the target through appearance, articulatory motion, or co-speech behavior. A still face mainly supplies identity information, synchronized lip motions additionally provides time-varying phoneme and viseme evidence, and body gestures provide a broader behavioral cue when the face is small or partially unavailable~\cite{pan2022seg}.
These cues are complementary rather than interchangeable: appearance identifies the target speaker, while synchronized motion captures articulation dynamics and temporal correspondence. They therefore require different levels of camera coverage, temporal alignment, and visual resolution.

Static visual cues provide speaker identity without requiring temporal synchronization. 
FaceFilter~\cite{chung2020facefilter} uses a single face image and a pretrained face-voice identity model to derive visual conditioning features for target-mask estimation. Qu \emph{et al.}~\cite{qu2020voiceface} further explores the complementarity between face and voice embeddings by conditioning extraction on pretrained face-recognition representations. 
These approaches demonstrate that facial appearance can specify the target  without synchronized visual streams, but they mainly provide identity information and cannot capture speech articulation dynamics.

To exploit fine-grained audio-visual correspondence, subsequent audio-visual TSE (AV-TSE) systems focus on synchronized lip movements. Early audio-visual separation models, such as Looking to Listen~\cite{ephrat2018looking} and VisualVoice~\cite{gao2021visualvoice}, establish general audio-visual fusion mechanisms, while direct AV-TSE methods incorporate visual cues into target-conditioned extraction. Wu \emph{et al.}~\cite{wu2019time} extends TasNet into an audio-visual time-domain extractor (AV-ConvTasNet), where frame-level lip representations are aligned with mixture features to estimate the target signal. MuSE~\cite{pan2021muse} further introduces self-enrolled speaker representations from intermediate estimates, combining visual articulation cues with speaker identity information without requiring a reference utterance.

Following this direction, later methods investigate richer visual representations and more robust conditioning strategies. Reentry~\cite{pan2021reentry} adopts pretrained audio-visual synchronization representations, while USEV~\cite{pan2022usev} explores alternative separator architectures. LiMuSE~\cite{liu2023limuse} additionally incorporates voiceprint  to complement visual cues. DAVSE~\cite{li2023rethinking} explicitly disentangles speaker identity and synchronization information, showing that both factors contribute to extraction performance. VCSE~\cite{li2022vcse} introduces a two-stage framework that first exploits visual cues for coarse extraction and then refines the result using self-enrolled phonetic representations.

Beyond cue representation, robustness to realistic visual conditions has also attracted increasing attention. PIAVE~\cite{liu2023piave} addresses pose variations by normalizing non-frontal faces and combining normalized and original observations, improving robustness under single-camera settings.

\subsubsection{Conditioning and Fusion}

Multimodal AV-TSE combines visual observations with enrollment audio to exploit complementary speaker cues. SpeakerBeam~\cite{ochiai2019multimodal} establishes an early fusion by combining global speaker representations from enrollment speech with frame-level visual features through attention. It optimizes audio-visual, audio-only, and visual-only extraction objectives, demonstrating that speaker identity and visible articulation provide complementary target information.

Subsequent AV-TSE studies move beyond convolutional neural network (CNN)-based fusion toward attention-driven architectures.

AV-SepFormer~\cite{Lin2023sepformer} uses cross-modal Transformers to extract target-relevant acoustic features from lips, while SAV-GridNet~\cite{pan2023scenario} and ImagineNet~\cite{pan2023imaginenet} introduce scenario-aware modeling and audio-visual correspondence learning, respectively. UniNet~\cite{wu2024unified} separates enrollment audio and lip cues into dedicated branches, and AV-CrossNet~\cite{kalkhorani2025avcrossnet} further explores unified multi-task modeling with early fusion and multi-level attention.

Since audio and visual cues have different reliabilities under real-world conditions, adaptive fusion has become an important direction. Multimodal attention fusion~\cite{sato2021multimodal} learns cue weights according to input reliability, while modality-dropout training~\cite{korse2024modality} improves robustness by preventing dependence on a single modality. These approaches differ from fixed fusion schemes by allowing cue contributions to vary with the observed conditions.

Recent methods increasingly leverage pretrained audio-visual representations. AV-HuMAR~\cite{wu2024avmar} adopts AV-HuBERT features and introduces Mask-and-Recovery training to exploit visual structure and audio-visual correspondence. AVSepChain~\cite{mu2024speechchain} decomposes extraction into speech perception and reconstruction stages, where lip representations and extracted speech mutually guide each other. The model-agnostic $C^{2}$AV-TSE framework~\cite{wu2025c} further introduces contextual training, confidence-guided refinement, and mask-recovery fine-tuning. Linguistic supervision from pretrained speech-language models provides additional constraints without affecting inference complexity~\cite{wu2025linguistic}. SEANet~\cite{tao2025audio} explores a complementary perspective by explicitly estimating interference through reverse selective auditory attention guided by visual cues.

Beyond lip motion, AV-TSE also incorporates richer visual context. Dual-path TSE~\cite{xu2023dual} extracts face representations using FaceNet and performs audio-visual fusion through dual-path cross attention. DCCRN-MultiClue~\cite{li2023clue} integrates video, textual, and sound-tag cues into a unified embedding space, while co-occurring-face attention~\cite{pan2025cooccurring} exploits other visible speakers as contextual activity cues. SEG~\cite{pan2022seg} extends visual conditioning from facial motion to co-speech gestures by using upper-body pose sequences as references. SeLG~\cite{pan2026beyondlips} further combines lip motion and gestures through cross-attention and contrastive alignment, demonstrating their complementary roles under complete and missing-modality conditions.

Beyond offline extraction, recent studies address dynamic and causal scenarios. ActiveExtract~\cite{li2024activeextract} incorporates active-speaker detection features to exploit speaking activity and synchronization information in sparsely overlapped conversations. MoMuSE~\cite{li2024momuse} maintains temporal speaker identity through momentum-based memory, improving robustness when visual observations are unreliable. For online extraction, 2S-AVTSE~\cite{twostagetse}, AV-ASEN~\cite{avase}, and Online AV-CrossNet~\cite{yu2025onlineavcrossnet} introduce lightweight visual encoders, causal architectures, and compressed spectral models to support real-time processing.

\subsubsection{Practical Limitations}
Visual conditioning depends on camera coverage and reliable tracking of the target. Occlusion, limited field of view, low resolution, illumination changes, and face-tracking errors can make the cue missing or misleading. Lip-based systems additionally require accurate audio-video synchronization, while still-face systems lack direct evidence of speech activity. Systems that combine visual cues with enrollment audio must also avoid relying on a corrupted modality when the other cue remains informative~\cite{sato2021multimodal,korse2024modality}. Memory-based conditioning and autoregressive acoustic cues can bridge short visual interruptions, but they introduce the additional problem of updating the stored identity when attention shifts~\cite{li2024momuse,avase}. Confidence-aware refinement addresses unreliable output segments, while co-occurring-face conditioning requires robust association between face tracks and acoustic sources~\cite{wu2025c,pan2025cooccurring}. These limitations motivate explicit handling of missing, asynchronous, or incorrectly associated observations.

%% file: sections/03_3_spatial_cues.tex
\subsection{Spatial Cues}

Spatially conditioned TSE exploits the spatial relation between the target and the recording device. Existing approaches can be grouped into three categories: spatial filtering-based methods, explicit spatial-query conditioning, and joint speaker-spatial modeling.

\subsubsection{Spatial Filtering-Based TSE}

Early spatially guided systems combine neural extraction with beamforming. Direction-Aware SpeakerBeam~\cite{li2019directionaware} selects target-related beam outputs using enrollment information, while NSF~\cite{gu2019neural} incorporates directional power-ratio and signal-to-noise features derived from beamformers together with spectral and phase features. Subsequent methods gradually replace fixed spatial processing with learnable filtering. Channel-decorrelation TSE extracts inter-channel differential information for time-domain extraction~\cite{han2021multichannel}, cNSF extends neural spatial filtering into the complex domain with optional MVDR reconstruction~\cite{gu2021complex}, and AN-BF learns beamforming refinement from neural estimates~\cite{gu2022towards}. Other approaches, including BG-TSE and iCOSPA, exploit target-steered beam outputs or location-aware spatial autoencoders~\cite{elminshawi2023beamformer,briegleb2023icospa}. Although these methods differ in the degree of integration between beamforming and neural extraction, they share the goal of transforming multichannel observations into target-focused representations.

\subsubsection{Explicit Spatial Query Conditioning}

Rather than relying on implicit spatial filtering, another family directly provides a spatial query to guide extraction. 3DSF~\cite{gu20213d} represents the target using azimuth, elevation, and distance, while DoA-assisted TSE~\cite{wang2024study} studies robustness to DoA uncertainty through knowledge distillation. Direction-based BiTSE~\cite{wang2025leveraging} and Location-Aware TSE~\cite{alcalapadilla2025location} explore compact direction representations, whereas distance-based TSE~\cite{shi2025distance} conditions extraction on a requested distance threshold.

Spatial queries can also define a target region rather than a specific speaker. TS-TSE~\cite{xu2022learning}, ReZero~\cite{gu2024rezero}, BASNet~\cite{yang2024binaural}, and M2M-TSE~\cite{choi2025multichannel} extract sources satisfying directional or temporal constraints, extending TSE from speaker selection toward spatially controlled source extraction. End-to-end DoA-guided extraction~\cite{jing2025doaguided} further represents the target region using direction and beamwidth embeddings, enabling neural beam selection without explicit beamforming.

\subsubsection{Joint Speaker and Spatial Conditioning}

Spatial cues are often combined with  identity cues when either modality is insufficient. Multi-modal multi-channel target speech separation jointly exploits target direction, enrollment speech, and lip motion through factorized attention~\cite{gu2020multi}. L-SpEx~\cite{ge2022spex} removes the assumption of known DoA by jointly estimating the target direction and extracting speech using enrollment-derived speaker representations. In contrast, NS-Extractor~\cite{lin2023focus} first obtains a spatially guided estimate and then derives a self-enrolled speaker representation for extraction. For moving speakers, weakly guided spatial filtering jointly learns tracking and extraction from an initial direction estimate~\cite{kienegger2025steering}.

Recent works further investigate adaptive fusion between speaker and spatial cues. Binaural TSE~\cite{meng2024binaural} combines speaker embeddings with binaural interaction features through attention-based filtering. MC-LExt~\cite{ling2026mclext} jointly learns speaker and spatial representations by prepending enrollment speech to multichannel mixtures without explicit DoA input. Eisenberg \emph{et al.}~\cite{eisenberg2026spectralspatial} introduce reliability-aware fusion to select informative cues under corrupted conditions. SSDQ~\cite{zhu2025ssdq} extends spatial conditioning to multimodal queries by jointly encoding semantic descriptions and spatial regions, enabling extraction under ambiguous target specifications.

\subsubsection{Spatial-Cue Encoding}

Spatial conditioning relies on two complementary representations: mixture-derived spatial observations and explicit target queries. Spatial observations characterize inter-channel relationships through features such as IPD, ITD, time difference of arrival (TDOA), spatial correlation, and beamformer outputs. NSF combines phase and spectral features with directional beamformer representations~\cite{gu2019neural}, while cNSF extends spatial features into the complex domain~\cite{gu2021complex}. DoA-assisted TSE further improves phase representation by reducing discontinuities caused by phase wrapping~\cite{wang2024study}.

Target queries should be encoded into forms that can be interpreted by neural models. Direction can be encoded using scalar values, categorical vectors, sinusoidal embeddings, or complex-exponential representations. Boolean directivity embedding provides a binary angular representation~\cite{wang2025leveraging}, while location-aware TSE encodes azimuth using complex exponential components~\cite{alcalapadilla2025location}. Region- and distance-based methods aggregate spatial samples or learn continuous distance embeddings to represent broader target conditions~\cite{gu2024rezero,shi2025distance}. The appropriate representation depends on whether the query specifies a point location, a spatial region, or a time-varying trajectory.

Beyond explicit spatial features, all-neural directional extraction~\cite{pandey2024directional} learns channel- and frame-level spatial representations directly from discretized azimuth and elevation. By injecting frame-level DoA embeddings into a causal extractor, such methods support dynamic target selection and reduce dependence on fixed utterance-level spatial queries.

\subsubsection{Practical Limitations}
Spatial cues remove the need for prerecorded identity references but rely on reliable spatial separation. Ambiguities arise when speakers overlap spatially or when location queries change with speaker motion. Their robustness is further limited by phase wrapping, array geometry, localization errors, and room-dependent distance estimation~\cite{wang2024study,lin2023focus}. Although region-conditioned systems provide flexible spatial control, they select targets by location rather than identity. Spatial cues therefore benefit most from reliable localization or complementary identity cues.

%% file: sections/03_4_textual_cues.tex
\subsection{Textual or Semantic Cues}
Textual or semantic conditioning specifies the target through descriptions, attributes, content, relations, or conversational context. This subsection reviews the corresponding language representations and fusion strategies, distinguishes the principal cue types, and summarizes their training objectives and ambiguity-related limitations.

\subsubsection{Conditioning and Fusion}
Unlike audio or synchronized video, textual cues describe the target indirectly through concepts, attributes, content, relations, or conversational context. A text or audio-text encoder maps this information into a representation that conditions the acoustic separator, commonly through concatenation, feature-wise linear modulation (FiLM), gating, or attention.

Heterogeneous Target Speech Separation~\cite{tzinis2022heterogeneous} represents non-mutually-exclusive concepts such as loudness, gender, language, and spatial location and applies FiLM throughout a time-domain separator. Presentation-driven TSE~\cite{jiang2024ptse} instead uses limited and temporally unaligned presentation text. Its pTSE-T model encodes the prompt with a frozen CLAP text encoder, applies hierarchical FiLM to audio features, and uses a DPRNN mask estimator to recover the target waveform. The same study also proposes a text-speech retrieval formulation that first obtains candidate speech streams and then associates them with the presentation text through contrastive learning.

Free-form descriptions support more flexible speaker selection. LLM-TSE~\cite{hao2025typing} uses a LoRA-adapted LLaMA~2 encoder for typed descriptions and an optional audio encoder for enrollment speech. After linear projection, the available cue embeddings are concatenated and supplied to a TD-SpeakerBeam-based extractor. The model supports text-only and joint text-audio conditioning and can interpret instructions to extract or suppress a specified speaker. StyleTSE~\cite{huo2025beyond} combines a SepFormer-based separator with a bi-modality cue network that can process speaking-style descriptions, enrollment audio, or both. PNTP-TSE~\cite{seki2025language} conditions a ResUNet30 separator on a fixed CLAP prompt embedding. ISRC-TSE~\cite{dai2025inter} uses a LoRA-adapted LLaMA~3.2 encoder and injects relative-cue representations through FiLM at two stages of a dual-path Transformer. CSE~\cite{kim2025contextual} takes a different approach by using previous dialogue turns as an implicit clue; its models can use textual history alone or combine it with enrollment speech.

TPEech~\cite{jiang2026tpeech} uses historical dialogue text as contextual evidence and introduces an echo-cue block for joint target extraction and noise suppression. SSDQ~\cite{zhu2025ssdq} provides a cross-cue alternative in which a natural-language description is paired with a spatial region. These systems extend text conditioning from standalone attributes or instructions to conversational and jointly constrained target definitions.

Several adjacent target sound extraction systems provide relevant conditioning mechanisms. AudioSep~\cite{liu2024separate} injects a frozen CLAP text embedding through layer-wise FiLM. CLAPSep~\cite{ma2024clapsep} reuses pretrained CLAP text and audio encoders, applies FiLM to layer-wise audio features, and is trained with SDR and SI-SDR objectives. DPM-TSE~\cite{hai2024dpm} and SoloAudio~\cite{wang2025soloaudio} show how class or language conditions can guide diffusion in spectrogram or latent spaces. Waveformer~\cite{veluri2023real} demonstrates causal query conditioning for real-time target sound extraction. These systems are included for their transferable text encoding, fusion, and generative formulations; their target sound results are not direct evidence of TSE performance.

\subsubsection{Cue Types}
Closed-set systems use one-hot or multi-hot labels for attributes such as gender, language, loudness, or spatial proximity. Such labels support controlled training and evaluation but are limited to a predefined inventory and annotation scheme. Free-form systems instead encode captions, phrases, or instructions. AudioSep and CLAPSep use general audio-text embeddings for source descriptions, while LLM-TSE encodes natural-language directives for speaker extraction or suppression.

Text may also refer to information beyond a source name. pTSE-T uses slide or summary text as content-level context for presentation speech. StyleTSE describes how a person speaks, and PNTP-TSE uses paralinguistic and non-linguistic descriptions such as pitch, speaking rate, loudness, emotion, gender, and perceptual impression words. These are absolute descriptions because their interpretation does not explicitly require comparison with the interferer. ISRC-TSE instead uses inter-speaker relations, such as selecting the faster or higher-pitched speaker, making the cue dependent on the speakers present in the mixture. CSE and TPEech use dialogue history rather than a standalone target description. Together, these works distinguish attribute-based, content-based, relational, contextual, and jointly semantic-spatial cues.

\subsubsection{Cue-Aware Training Objectives}
Most text-conditioned TSE systems use language to condition a separator trained with a signal-reconstruction objective. pTSE-T~\cite{jiang2024ptse} and ISRC-TSE~\cite{dai2025inter} optimize negative SI-SDR, while the PNTP-TSE~\cite{seki2025language} baseline uses a waveform L1 loss. Heterogeneous Target Speech Separation predicts target and residual components under a mixture-consistency constraint~\cite{tzinis2022heterogeneous}. CSE illustrates a cue-dependent auxiliary objective: ContSep~\cite{kim2025contextual} combines permutation-invariant reconstruction with cross-entropy classification of the context-matched output stream, whereas ContExt directly reconstructs only the context-selected target. A contrastively pretrained encoder does not necessarily imply a contrastive separation objective. PNTP-TSE~\cite{seki2025language} keeps CLAP fixed, and the adjacent CLAPSep~\cite{ma2024clapsep} target sound system trains its separator with SDR and SI-SDR objectives.

\subsubsection{Practical Limitations}
Textual or semantic cues can identify a target without a prerecorded enrollment utterance, but their discriminative power depends on the scene. An absolute attribute can be shared by several speakers, a speaking-style attribute may vary within an utterance, and a relational cue requires the relevant contrast to be present in the mixture. Presentation text may describe content shared across speakers, while dialogue history may be insufficient to identify the next speaker. Free-form descriptions also introduce linguistic variation and possible mismatch between the concepts represented by the text encoder and the acoustic evidence available in the mixture. These limitations motivate ambiguity-aware prompting, combinations of complementary attributes, and optional fusion with enrollment or contextual cues~\cite{hao2025typing,huo2025beyond,dai2025inter,kim2025contextual}.

%% file: sections/03_5_neural_cues.tex
\subsection{Neural Cues}

Neural-conditioned TSE exploits physiological signals associated with selective auditory attention to identify the attended speech stream. This section reviews neural representations, conditioning strategies, and the challenges caused by listener and session variability.

\subsubsection{Cue Information and Encoding}

Neural signals provide information about the speech stream attended by a listener. Selective auditory attention has been investigated using electrocorticography (ECoG)~\cite{mesgarani2012selective}, magnetoencephalography (MEG)~\cite{ding2012emergence}, and electroencephalography (EEG)~\cite{o2015attentional,biesmans2016auditory}. By reconstructing the attended speech envelope from neural recordings and comparing it with candidate streams, these studies establish the foundation for neural-guided target selection~\cite{o2017neural,aroudi2020cognitive}.

\subsubsection{Conditioning and Fusion}

Early neural-guided extraction systems adopt a two-stage strategy: first separating the mixture into candidate streams and then selecting the stream that best matches the neural response~\cite{o2017neural,han2019speaker}. Although effective, this pipeline requires multiple separation outputs and does not jointly optimize extraction and neural conditioning.

Recent methods instead integrate neural cues into end-to-end extraction models. BESD~\cite{hosseini2021} fuses EEG and audio representations before recurrent mask estimation, while NeuroHeed~\cite{pan2023neuroheed} introduces a neural attractor derived from EEG and further explores online self-enrollment from previously extracted speech. NeuroSpEx~\cite{neurospex} employs cross-attention between EEG and speech representations, and M3ANet~\cite{m3anet} introduces contrastive temporal alignment with multiscale Mamba-based speech encoding. Beyond feature fusion, MLAD-ETSEN~\cite{wang2025mlad} jointly performs extraction and auditory-attention decoding through symmetric cross-attention between EEG and speech branches. TIDENet~\cite{lv2026tidenet} addresses EEG-speech misalignment through trainable interpolation and dual-path representation learning, while BM-TSE~\cite{han2026brainprint} learns personalized brain representations to improve subject adaptation and cross-session robustness. Recent work further combines EEG with eye tracking to capture dynamic attention changes during extraction~\cite{wang2026eegeye}.

Neural cues are incorporated through both reconstruction objectives and attention-related constraints. NeuroHeed~\cite{pan2023neuroheed} optimizes extraction using SI-SDR, while NeuroHeed+~\cite{pan2024neuroheed+} additionally introduces auditory-attention classification. NeuroSpEx+~\cite{neurospex+} incorporates attended-speech-envelope reconstruction and correlation-based objectives, and M3ANet~\cite{m3anet} uses contrastive learning to align neural and acoustic representations. These objectives encourage the extracted signal to remain consistent with the listener's attended stream.

\subsubsection{Practical Limitations}

Neural cues reflect the listener's attentional state, but EEG-based signals are noisy and exhibit substantial variability across subjects and recording sessions. Consequently, many existing systems rely on subject-dependent settings, and generalization to unseen listeners and environments remains a key challenge~\cite{pan2023neuroheed,neurospex,m3anet}.

%% file: sections/03_6_crossmodal_comparison.tex
\subsection{Cross-Cue Comparison}
The five cue categories can be compared along three axes: the evidence used to specify the target, its temporal specificity, and its acquisition requirements. Audio enrollment and still-face cues primarily provide persistent identity information, whereas synchronized lip motion and neural activity provide time-varying evidence about articulation or listener attention. Spatial cues select sources by direction, distance, or region, while textual cues express attributes, content, relations, or conversational context. The latter two may specify the desired source without persistent identity, allowing the selected speaker to change with the scene or instruction. Elminshawi \emph{et al.}~\cite{elminshawi2024auxiliary} compare audio- and video-conditioned extraction with uninformed separation under a common DPRNN backbone, showing that auxiliary cues do not uniformly improve extraction and can degrade under cue distortion. This highlights the need to distinguish cue effectiveness from improvements in the extraction backbone.

The cues also differ substantially in reliability and practical usability. Enrollment audio may be unavailable, corrupted, or insufficiently discriminative, while also requiring storage or processing of speaker-specific information with potential privacy concerns. Visual cues require camera access and, for lip-based systems, temporal synchronization, raising additional privacy and sensing requirements. Spatial cues avoid explicit identity information but require suitable microphone configurations, spatial separation, and accurate localization. Text is inexpensive to acquire and computationally convenient, but may be ambiguous or inconsistent with the acoustic scene. Neural cues directly reflect listener attention, yet current EEG-based systems suffer from noisy and subject-dependent signals and impose substantial sensing, calibration, power, and wearability burdens; invasive neural interfaces further raise practical and privacy concerns. These trade-offs are particularly important for hearing aids and wearable devices, where sensing hardware, device weight, battery life, and user comfort constrain deployment. No cue is therefore uniformly preferable: its suitability depends not only on informativeness and reliability, but also on privacy, sensing burden, and deployment constraints.

These differences make the cue categories complementary. Multi-modal multi-channel TSE combines direction, enrollment speech, and lip movement~\cite{gu2020multi}; multimodal attention and modality-dropout systems adapt to changes in audio and visual reliability~\cite{sato2021multimodal,korse2024modality}; SeLG combines lip and gesture evidence~\cite{pan2026beyondlips}; and textual systems combine descriptions or dialogue context with enrollment speech~\cite{hao2025typing,huo2025beyond,kim2025contextual}. Reliability-aware spectral-spatial fusion handles conflicting cues~\cite{eisenberg2026spectralspatial}, while EEG and eye tracking provide complementary measurements of listener attention~\cite{wang2026eegeye}. However, additional modalities also increase sensing, alignment, computation, and power requirements. For real-time applications such as hearing aids, the benefit of multimodal conditioning must therefore be balanced against latency, computational complexity, and energy consumption. Auxiliary objectives remain method-specific, including speaker discrimination, cross-modal alignment, cue-validity prediction, and semantic classification, while generative denoising, flow matching, and token prediction are discussed separately in Sec.~\ref{sec:evolution}.

%% file: sections/04_network_evolution.tex
\section{Evolution of TSE Networks 
}
\label{sec:evolution}

\begin{figure*}[!t]
  \centering
  \includegraphics[width=0.99\linewidth]{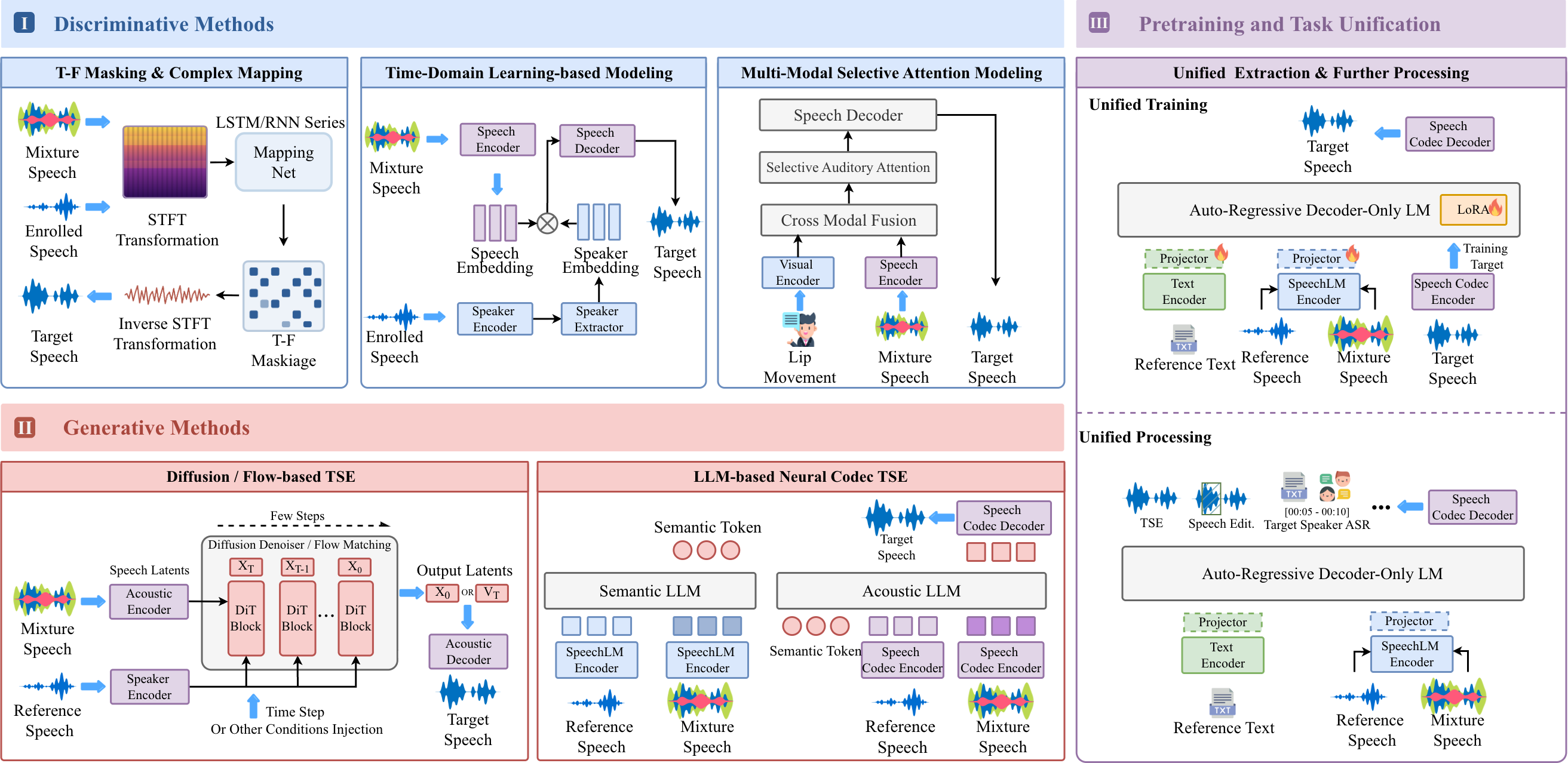}
      \caption{\textbf{Architecture of TSE networks.} Discriminative systems directly estimate a \ac{TF} representation or waveform of the target. The generative panel illustrates recent pipelines based on diffusion, flow matching, and LLMs. Earlier variational source modeling is reviewed in the text. The panel on the right presents pretraining and task unification as design dimensions that apply across paradigms. These approaches reuse large encoders, generative backbones, or shared models for TSE and related speech processing tasks. The cue encoders shown in each pipeline can be adapted to the auxiliary modalities reviewed in Sec.~\ref{sec:type_cues}.  }
  \label{fig:evolved_arch}
\end{figure*}

Sec.~\ref{sec:type_cues} categorizes TSE systems by target cues. In this section, we distinguish two primary estimation paradigms: direct discriminative estimation and probabilistic or conditional generative modeling. Hybrid systems combine components from both paradigms. Pretraining, foundation model adaptation, and task unification are design dimensions that apply across both paradigms. Fig.~\ref{fig:evolved_arch} summarizes these architectural trends. Together, these dimensions describe how targets are estimated and how model knowledge is reused.

\subsection{Discriminative Estimation}

Discriminative TSE systems estimate the target signal or its latent representation directly from $(\mathbf{x},\mathbf{Q}_{t})$. Their objectives generally follow the reconstruction and auxiliary-loss formulation in Eq.~\eqref{eq:generic_tse_loss}. Architectural evolution has been shaped by four key aspects: signal representation, sequence modeling, cue fusion, and causal deployment.

\subsubsection{Time-Frequency Masking and Mapping}

Early neural TSE systems commonly transformed the mixture into a \ac{STFT} representation and estimated a target-dependent \ac{TF} mask. SpeakerBeam used an auxiliary network to derive a speaker representation that adapted the extraction network~\cite{speakerbeam}, while VoiceFilter conditioned a spectrogram-mask estimator on a separately trained speaker embedding~\cite{wang2019voicefilter}. These systems established the encoder, conditioning, and extractor pattern that remains common in later TSE architectures.

\ac{TF} processing does not necessarily imply magnitude-only estimation. Later systems estimate complex masks or directly map mixture features to a complex target spectrum, allowing the network to modify both magnitude and phase. DPCCN uses densely connected complex-valued processing for robust separation and extraction~\cite{DPCNN}, while X-TF-GridNet combines \ac{TF} grid modeling with adaptive speaker-embedding fusion~\cite{XTFGRIDNET}. Thus, the phase-reconstruction limitation commonly attributed to \ac{TF} methods applies specifically to those that reuse the unmodified mixture phase, rather than to all \ac{TF}-based TSE systems.

CIENet exchanges contextual information between enrollment and mixture representations in the \ac{TF} domain, and a later coarse-to-fine framework performs target extraction at successive resolutions~\cite{cienet,yang2024coarsetofine}. These systems illustrate that \ac{TF} modeling can retain structured enrollment information rather than reducing the cue to one vector before separation.

\subsubsection{Time-Domain Waveform Estimation}

The success of learned waveform encoders in speech separation motivated end-to-end time-domain TSE. Conv-TasNet replaces the fixed STFT with a trainable one-dimensional encoder and decoder and estimates masks in the learned representation~\cite{convtasnet}. SpEx adapts this strategy to extraction by jointly learning the speaker representation and a multi-scale time-domain extraction network~\cite{spex}. SpEx+~\cite{spex+} further aligns the mixture and enrollment encoders through a shared speaker encoder, and SpEx++ introduces multi-stage refinement using utterance- and frame-level reference information~\cite{spex++}. 
These models avoid explicit mixture-phase reconstruction, but their performance still depends on the learned analysis basis, separator capacity, and target-conditioning mechanism.

\subsubsection{Attention and Structured Sequence Modeling}

Newer TSE architectures complement recurrent and temporal convolutional backbones with Transformers that use processing along two paths, Conformers, and models that operate on the \ac{TF} grid. X-SepFormer uses Transformer processing along two paths and training designed to reduce speaker confusion~\cite{xsepformer}. Systems based on Conformers combine local convolutional modeling with attention over longer contexts~\cite{sinha2022conformer}. SEF-Net removes the need for speaker embeddings extracted in advance through Conformer modeling within and across chunks~\cite{SEFNET}. Recent frameworks also improve target conditioning through fusion based on attractors, hierarchical speaker representations, attention between cue and mixture streams, and feature modulation~\cite{chetupalli2024unified,he2024hierarchical,mltse,usef}. These mechanisms provide a general interface for acoustic, visual, spatial, semantic, and neural cues. Designs for each modality are discussed in Sec.~\ref{sec:type_cues}.

\subsubsection{Pretrained Representations for Discriminative TSE}

Pretraining first influenced discriminative TSE by providing stronger representations of the mixture and target speaker. SSL-MHFA uses representations from a pretrained WavLM model to enhance mixture features and obtain an embedding of the target speaker before extraction~\cite{sslmhfa}. The downstream extractor still performs direct signal estimation. This use of pretraining therefore differs from foundation model adaptation, where the pretrained backbone itself is adapted as the conditional restoration or generation model.

\subsubsection{Causal and Low-Latency Networks}

Streaming TSE requires balancing extraction quality with  computational cost, as future mixture frames are unavailable during causal inference. Recent studies address this challenge through causal architectures and efficient target conditioning. UniSEF investigates global and low-latency time-domain variants based on dual-path recurrent processing and analyzes the trade-off between latency and extraction performance~\cite{uninet}. Personalized PercepNet enables causal low-complexity enhancement by conditioning a compact spectral model on target-speaker embeddings~\cite{personalized}, while DSINet~\cite{hao2025dsinet}, TargetVoice~\cite{pallala2025targetvoice}, and SpeakerBeam-SS~\cite{speakerbeamss} further reduce runtime cost through dynamic speaker fusion, compact edge deployment, and efficient state-space modeling. These advances highlight that practical streaming TSE requires joint optimization of latency, efficiency, and extraction quality.

\subsection{Probabilistic and Conditional Generative Modeling}

Rather than relying solely on direct point estimates, these methods model the target or interference speech with explicit probabilistic or generative components. They include probabilistic source models and conditional generators based on diffusion, flow matching, or token prediction. To describe continuous generative objectives across signal and latent spaces, we use $\mathbf{r}_{t}=h(\mathbf{s}_{t})$ to denote the target representation modeled by a particular system. The transform $h(\cdot)$ may be the identity transform, an \ac{STFT} or mel transform, or a learned latent encoder. The following objectives provide schematic descriptions of these model families. Individual methods may use different formulations.

\subsubsection{Variational Generative Source Modeling}

Published probabilistic TSE predates recent formulations based on diffusion. An early system uses two channels and combines geometric constraints derived from the target direction with separate conditional variational autoencoders for the target speech and the interference mixture. A \ac{TF} mask is then used for postprocessing~\cite{wang2022directioncvae}. A subsequent extension replaces fixed source labels with global style tokens so that the CVAE conditions can be inferred from noisy observations~\cite{wang2023directioncvae}. A later study modifies the optimization to reduce sensitivity to errors in the target direction~\cite{wang2024dualcvae}. Another system jointly trains a target model conditioned on global style tokens and a neural postfilter for noisy underdetermined conditions~\cite{wang2025cvaepostfilter}.

These systems form hybrid probabilistic separation pipelines rather than waveform generators initialized from noise. Directional information selects the target. CVAEs provide learned spectrogram source models within geometric source separation. Masking or neural postfiltering then produces the final estimate. These systems therefore form an earlier line of generative source models that complements methods based on diffusion, flow matching, and token generation.

\subsubsection{Conditional Diffusion}

Diffusion models for TSE corrupt the target representation through a forward process and train a conditional denoiser or score model to reverse that process. A representative objective for noise prediction can be written as
\begin{equation}\label{eq:diff}
    \mathcal{L}_{\mathrm{diff}}
    =
    \mathbb{E}_{\substack{
    (\mathbf{r}_{t},\mathbf{x},\mathbf{Q}_{t})\sim p_{\mathrm{data}}\\
    \tau\sim\pi_{\mathrm{diff}},\,
    \boldsymbol{\epsilon}\sim\mathcal{N}(\mathbf{0},\mathbf{I})}}
    \left[
    \left\|
    \boldsymbol{\epsilon}
    -
    \boldsymbol{\epsilon}_{\theta}
    \left(
    \mathbf{r}_{t}^{(\tau)},\tau,\mathbf{x},\mathbf{Q}_{t}
    \right)
    \right\|_{2}^{2}
    \right],
\end{equation}
where $p_{\mathrm{data}}$ is the empirical distribution of training tuples and $\pi_{\mathrm{diff}}$ is the sampling distribution over normalized diffusion time $\tau\in[0,1]$. At time $\tau$, the perturbed target representation is $\mathbf{r}_{t}^{(\tau)}=\gamma_{\tau}\mathbf{r}_{t}+\sigma_{\tau}\boldsymbol{\epsilon}$. The coefficients $\gamma_{\tau}$ and $\sigma_{\tau}$ control the signal and noise levels according to the chosen noise schedule. The matrix $\mathbf{I}$ is the identity covariance matrix with the required dimension. The function $\boldsymbol{\epsilon}_{\theta}(\cdot)$ is the conditional noise predictor parameterized by $\theta$. Here, the superscript in parentheses indexes diffusion time. Under this perturbation parameterization, the corresponding conditional score estimate can be written as
\[
\boldsymbol{\psi}_{\theta}
\left(
\mathbf{r}_{t}^{(\tau)},\tau,\mathbf{x},\mathbf{Q}_{t}
\right)
=
-\frac{
\boldsymbol{\epsilon}_{\theta}
\left(
\mathbf{r}_{t}^{(\tau)},\tau,\mathbf{x},\mathbf{Q}_{t}
\right)
}{\sigma_{\tau}}.
\]
Diff-TSE conditions the diffusion model on the mixture and a cue for the target speaker. It investigates alternative configurations of the score model and ensemble inference~\cite{difftse}. DDTSE combines a diffusion forward process with a discriminative reconstruction objective and a training procedure with two stages that reproduces the inference procedure during training. This design allows fewer reverse steps and can refine a discriminative estimate~\cite{ddtse}. These systems show that diffusion and direct estimation can be combined rather than treated as strictly separate families. In the related task of target sound extraction, DPM-TSE uses diffusion in the spectrogram domain conditioned on sound classes~\cite{hai2024dpm}. SoloAudio uses latent diffusion conditioned on language~\cite{wang2025soloaudio}. These methods are included as examples of transferable conditioning formulations rather than direct evidence for TSE.

\subsubsection{Flow Matching}

Conditional flow matching learns a velocity field that transports an initial distribution $p_{0}$ to the conditional distribution of the target representation $\mathbf{r}_{t}$ given $\mathbf{x}$ and $\mathbf{Q}_{t}$. For $\mathbf{z}_{0}\sim p_{0}$ and $\mathbf{z}_{1}=\mathbf{r}_{t}$, let $\mathbf{z}_{\tau}$ denote the state at normalized time $\tau\in[0,1]$ along the chosen path. A generic objective is
\begin{equation}\label{eq:fm}
    \mathcal{L}_{\mathrm{FM}}
    =
    \mathbb{E}
    \left[
    \left\|
    \mathbf{v}_{\theta}
    (\mathbf{z}_{\tau},\tau,\mathbf{x},\mathbf{Q}_{t})
    -
    \mathbf{u}_{\tau}
    \right\|_{2}^{2}
    \right],
\end{equation}
where $\mathbf{v}_{\theta}(\cdot)$ is the learned conditional velocity field and $\mathbf{u}_{\tau}$ is the target velocity of the chosen path. The expectation is taken over training tuples, initial states, sampled times, and the corresponding path states. The initial state may be noise or a deterministic estimate derived from the mixture or background. During inference, the learned velocity field is integrated to recover the target representation.

FlowTSE applies conditional flow matching in the mel spectrogram domain and uses a vocoder conditioned on the mixture \ac{STFT} for phase reconstruction~\cite{flowtse}. Whisper-FMTSE adds semantic supervision from a pretrained ASR encoder to improve intelligibility~\cite{fmtse}. AD-FlowTSE defines a deterministic path from the background to the target and adjusts the initialization and integration interval using an estimated mixing ratio~\cite{hsieh2025adaptive}. These systems mainly differ in their representation space, initialization, path construction, and numerical solver.

\subsubsection{Discrete Speech, Neural Codecs, and Language Models}

Another route converts the target into discrete speech units or tokens from a neural codec and predicts these representations before waveform synthesis. Let $\mathbf{y}_{1:L}$ denote the token sequence obtained by applying a speech tokenizer or neural codec to the target speech $\mathbf{s}_{t}$. A typical autoregressive objective is
\begin{equation}\label{eq:tok}
    \mathcal{L}_{\mathrm{tok}}
    =
    -\sum_{\ell=1}^{L}
    \log p_{\theta}
    \left(
    \mathbf{y}_{\ell}
    \mid
    \mathbf{y}_{<\ell},\mathbf{x},\mathbf{Q}_{t}
    \right),
\end{equation}
where $L$ is the sequence length, $\mathbf{y}_{<\ell}$ collects the previously predicted tokens, and $p_{\theta}$ denotes the conditional token distribution parameterized by $\theta$. With a tokenizer that uses one codebook, $\mathbf{y}_{\ell}$ contains one token index. With a residual codec or a codec that uses multiple codebooks, it contains the vector of tokens from all retained codebooks at step $\ell$. Thus, $L$ reflects the token rate rather than the waveform sampling rate. A system based on speech discretization predicts target speech units from the mixture and enrollment speech, then reconstructs the waveform with a separately trained discrete vocoder~\cite{skim}. TSELM discretizes representations from multiple WavLM layers. It uses attention between the mixture and target speaker representations, predicts target tokens with a language model, and reconstructs audio with a scalable HiFi-GAN~\cite{tselm}. LauraTSE uses continuous representations of the mixture and enrollment speech as input to an autoregressive language model with an architecture that contains only a decoder. It predicts tokens from the first layers of the target codec and reconstructs the complete codec representation with an additional model that contains only an encoder~\cite{lauratse}. Systems based on tokens introduce a representation bottleneck and a separate decoding stage. Their behavior therefore depends on token resolution, prediction errors, and the reconstruction ability of the codec or vocoder.

\subsection{Pretraining and Task Unification}

Pretraining and task unification apply across the discriminative and generative paradigms described above. Approaches based on foundation models transfer a generative or representational prior learned from large speech or audio corpora. The NeMo speech restoration model is pretrained through conditional flow matching directly on complex \ac{STFT} coefficients. It is then adapted to several restoration tasks, including TSE, without requiring a mel spectrogram vocoder~\cite{nemo}. Metis uses masked generative pretraining over speech tokens learned through self supervision and is adapted to several speech generation tasks, including TSE~\cite{wang2025metis}. UniAudio is a broader foundation model for audio generation that includes TSE among many token generation tasks~\cite{uniaudio}. These models demonstrate transfer across tasks. However, a unified model is not automatically specialized for errors in target selection, latency constraints, or the signal fidelity requirements of TSE.

Unified speech models place extraction within a shared collection of speech transformation tasks. SpeechX uses a language model over neural codec tokens for TSE, speech enhancement, speech removal, editing, and speech synthesis from text~\cite{speechx}. AnyEnhance uses masked generative modeling, guidance from a reference prompt, and an internal critic mechanism to support TSE together with several restoration tasks for speech and singing voice~\cite{anyenhance}. LLaSE-G1 combines continuous WavLM inputs, a predictor based on LLaMA, and targets produced by X-Codec2. Its formulation uses two channels and is evaluated on noise suppression, packet loss concealment, TSE, acoustic echo cancellation, and speech separation~\cite{llaseg1}. In this review, these systems are included when they explicitly evaluate TSE conditioned on enrollment speech or contribute a transferable conditioning or decoding mechanism. They are not treated as evidence that one architecture or objective is optimal for every TSE setting.

\subsection{Comparison of Network Paradigms}

The estimation paradigms emphasize different properties. Discriminative systems usually produce an estimate in one forward pass. They directly optimize reconstruction criteria based on reference signals and can support extraction with low latency. Variational spatial systems introduce learned source priors but retain assumptions about multichannel geometry and iterative optimization of source models. Diffusion and flow models introduce conditional denoising or transport trajectories. Their inference cost depends on the number of model evaluations and the chosen solver. Models based on discrete tokens operate at a lower temporal resolution than waveform generation but introduce quantization and codec reconstruction. Hybrid systems combine components from these families to balance reconstruction fidelity, robustness, and inference cost. Pretraining and task unification apply to all of these paradigms and can transfer representations across datasets and tasks. Model scale, adaptation strategy, and decoding pipelines may nevertheless complicate deployment. Network comparisons should therefore report signal fidelity, perceptual quality, intelligibility, speaker consistency, computational cost, and latency rather than assume that one paradigm is uniformly superior.

%% file: sections/05_datasets.tex
\section{Datasets and Benchmark Construction}
\label{sec:dataset}

Dataset choice in TSE affects the observed speakers, acoustic conditions, target cues, and evaluation protocols. We distinguish \emph{source corpora}, which provide speech or cue recordings, from \emph{TSE benchmarks}, which define mixtures, target assignments, auxiliary cues, and evaluation splits. Table~\ref{tab:dataset} and~\ref{tab:cue_dataset} summarize these resources.

\input{table_dataset}

\subsection{Source Speech and Enrollment Corpora}

WSJ0 and LibriSpeech provide clean read speech that is widely used to construct TSE mixtures and enrollment utterances~\cite{WSJ0,2015librispeech}. They are source corpora rather than TSE benchmarks: typical pipelines select utterances from multiple speakers, create mixtures at controlled signal-to-interference ratios, and reserve target-speaker recordings as enrollment cues. Speaker-disjoint splits are essential to evaluate generalization and avoid speaker-identity leakage.

VoxCeleb1 and VoxCeleb2 provide large-scale speaker-labeled speech collected from unconstrained online videos and are commonly used for speaker encoder training, enrollment-target pairing, and audio-visual source construction~\cite{Voxceleb,Voxceleb2}. However, natural source recordings do not necessarily imply natural acoustic mixtures, as many TSE studies still generate mixtures by digitally combining independent recordings. Other corpora, such as VCTK and AISHELL-1, provide additional diversity in accent, language, and recording conditions~\cite{CSTR,AISHELL-1}.

\subsection{Mixture and Acoustic-Scene Construction}

Most  TSE studies rely on simulated mixtures generated from separately recorded sources. WSJ0-2Mix and WSJ0-3Mix establish controlled benchmarks by combining clean utterances with known source signals~\cite{hershey2016deep}. LibriMix extends this paradigm using LibriSpeech to provide clean, noisy, two-speaker, and three-speaker mixtures, together with sparsely overlapping evaluation scenarios~\cite{librimix}. Since the original sources are available, these benchmarks support waveform-level supervision and reference-based metrics such as SI-SDR, while enabling controlled variation of speaker identity, overlap ratio, and signal level. However, independently sampled utterances cannot fully capture conversational dynamics, correlated background activity, or source motion in real interactions.

Specialized simulated datasets further evaluate specific robustness factors. wTIMIT2mix constructs mixtures and enrollment pairs from normal and whispered speech to study matched and mismatched speaking conditions~\cite{borsdorf2024wtimit2mix}. Noise, reverberation, and spatial characteristics are commonly added through additional simulation stages. WHAM! and WHAMR! introduce real noise and simulated reverberation into WSJ0 mixtures~\cite{WHAM!,WHAMR!}, while SMS-WSJ generates multichannel spatial mixtures using simulated rooms, microphone arrays, and source positions~\cite{smswsj}. Similar procedures are widely adopted in spatial TSE by combining room impulse responses (RIR) or head-related transfer functions (HRTF) with controlled source locations. Although these datasets provide accurate waveform and spatial supervision, their generalization depends on the fidelity of  simulation and the coverage of generation parameters.

Real-recorded mixtures provide greater acoustic realism. LibriCSS creates conversation-like recordings by replaying LibriSpeech utterances in real rooms and capturing them with microphone arrays under different overlap conditions~\cite{libricss}. AISHELL-4 provides natural Mandarin meeting recordings with speaker activity annotations and transcripts, while CHiME-5 and DiPCo capture far-field conversational speech using distributed arrays and close-talking microphones~\cite{Aishell-4,chime5,dipco}. REAL-T further converts natural conversations from AISHELL-4, AliMeeting, AMI, CHiME-6, and DiPCo into a TSE benchmark by pairing overlapping target segments with non-overlapping enrollment utterances~\cite{li2025realt}. However, these datasets generally lack isolated target signals at the microphone arrays, limiting their use for conventional waveform-level supervision. LOCATA provides controlled real recordings and source trajectories for spatial localization and tracking, but is not designed as a general TSE corpus~\cite{locata}.

This difference explains the dominance of simulated mixtures in TSE. Controlled synthesis enables exact target reconstruction and quantitative evaluation, whereas real recordings are primarily used for ASR-based evaluation, robustness analysis, qualitative assessment, or domain adaptation under appropriate protocols.

\subsection{Cue-Bearing and Multimodal Datasets}

Visual TSE typically derives cues from audio-visual speech corpora and constructs mixtures afterward. Controlled datasets, including GRID, CUAVE, TCD-TIMIT, and OuluVS2, provide synchronized facial or lip observations, while LRS2, LRS3, VoxCeleb2, and AVSpeech offer larger-scale in-the-wild recordings with diverse speakers, poses, and environments~\cite{cooke2006audio,CUAVEAN,TCD-TIMIT,OuluVS2,LRS2,afouras2018lrs3,Voxceleb2,AVSpeech}. AVA-ActiveSpeaker~\cite{AVA-ActiveSpeaker} provides visible-speaker annotations but is primarily designed for active-speaker detection rather than waveform extraction. For typical AV-TSE, target and interfering speech are mixed while retaining the target visual stream. Such datasets provide exact target supervision without capturing the acoustic-visual interaction of real multi-speaker scenes.

Text-conditioned TSE requires associating speakers or utterances with language descriptions. Existing approaches derive textual cues from metadata, transcripts, speaker attributes, or generated comparisons between mixture speakers. TextrolSpeech~\cite{ji2024textrolspeech} provides style-annotated speech, TextrolMix~\cite{huo2025beyond} associates mixtures with speaking-style descriptions, and PromptTSE explores para-linguistic and non-linguistic queries~\cite{seki2025language}. Since annotation strategies vary, textual cues may represent speaker attributes, inter-speaker relations, dialogue context, or speech content.

Some benchmarks combine multiple auxiliary cues. SS-Libri associates spatialized mixtures with semantic descriptions and region queries, enabling evaluation of joint semantic-spatial conditioning~\cite{zhu2025ssdq}. However, similar to most benchmarks, it does not represent naturally recorded conversations.

Neural-conditioned TSE relies on smaller datasets that synchronize speech stimuli with neural recordings. The KU Leuven dataset, BISS-EEG, and BISS-iEEG provide EEG or intracranial recordings under competing-speech conditions, while the Cocktail Party dataset captures neural responses to continuous speech narratives~\cite{das2019auditory,fuglsang2020effects,ieeg_data,broderick2018electrophysiological}. AVED and MM-AAD further incorporate visual observations with neural measurements~\cite{MSFNet,fan2025seeing}. These datasets preserve the original stimuli and neural responses, but their limited participant populations, recording sessions, and stimulus diversity remain major challenges for generalization.

\subsection{Benchmark Characteristics and Comparability}

Meaningful dataset comparison requires reporting several key properties. These include the role of the resource (\emph{e.g.}, source corpus, mixture benchmark, cue-bearing dataset, or real recording), mixture construction settings (overlap, noise, reverberation, and spatial conditions), target cue generation, availability of isolated target references, and speaker/session/source partitioning. These factors determine whether waveform-level supervision is possible and whether evaluation is affected by identity or content leakage.

Datasets derived from the same speech corpus are not necessarily comparable, as studies may differ in enrollment selection, mixture generation, spatial simulation, noise conditions, or target definitions. Tables~\ref{tab:dataset} and~\ref{tab:cue_dataset} therefore summarize how each resource is constructed and used rather than relying only on dataset scale. The effects of limited real-mixture references, inconsistent simulation protocols, and uneven modality coverage are discussed in Sec.~\ref{sec:challenge}.

Open-source recipes improve reproducibility by standardizing data construction and training pipelines. For example, WeSep provides structured TSE recipes, scalable mixture simulation, target-speaker modeling, and deployment support~\cite{wesep}. However, different configurations within the same toolkit may still produce non-equivalent benchmarks, making detailed recipe reporting essential.

%% file: table_dataset.tex
\begin{table*}[!t]
\caption{Representative source, mixture, augmentation, and recorded-scene resources used in TSE research. ``Exact reference'' indicates whether the target signal used to form the evaluated mixture is available for waveform supervision or signal-level scoring.}
\label{tab:dataset}
\centering
\footnotesize
\setlength{\tabcolsep}{2pt}
\renewcommand{\arraystretch}{1.04}
\begin{tabularx}{\textwidth}{@{}
>{\raggedright\arraybackslash}p{0.14\textwidth}
>{\raggedright\arraybackslash}p{0.145\textwidth}
>{\raggedright\arraybackslash}p{0.12\textwidth}
>{\raggedright\arraybackslash}p{0.255\textwidth}
>{\raggedright\arraybackslash}p{0.105\textwidth}
>{\raggedright\arraybackslash}X@{}}
\toprule
Resource & Primary role & Cue or modality & Mixture or recording construction & Exact reference & Typical use in TSE research \\
\midrule
\rowcolor{gray!20}\multicolumn{6}{@{}c}{\rule{0pt}{2.6ex}\small\textbf{Source speech and enrollment corpora}}\\[0.6ex]

WSJ0~\cite{WSJ0} & Clean source corpus & Audio identity & Separate read utterances; mixtures generated downstream & Not a mixture & Target, interference, and enrollment speech \\
LibriSpeech~\cite{2015librispeech} & Clean source corpus & Identity; transcript & Separate audiobook utterances; mixtures generated downstream & Not a mixture & Large-scale source and enrollment speech \\
VoxCeleb1/2~\cite{Voxceleb,Voxceleb2} & In-the-wild source corpus & Audio-visual identity & Natural video clips; TSE mixtures usually combine independent clips & Source known & Speaker encoders, enrollment, and face-conditioned TSE \\
VCTK and AISHELL-1~\cite{CSTR,AISHELL-1} & Source corpora & Accent; language & Separate utterances with downstream mixing and augmentation & Not a mixture & Cross-corpus, accent, and language evaluation \\
\midrule
\rowcolor{gray!20}\multicolumn{6}{@{}c}{\rule{0pt}{2.6ex}\small\textbf{Generated mixtures and augmentation resources}}\\[0.6ex]
WSJ0-2Mix/3Mix~\cite{hershey2016deep} & Clean mixture benchmark & Assigned audio identity & Digital mixing of independent WSJ0 speakers & Yes & Supervised training and signal-level comparison \\
LibriMix and SparseLibriMix~\cite{librimix} & Clean or noisy benchmark & Assigned audio identity & Digital two- or three-speaker mixing; optional noise and sparse overlap & Yes & Generalization, noise, and overlap evaluation \\
wTIMIT2mix~\cite{borsdorf2024wtimit2mix} & Speech-mode mixture benchmark & Enrollment identity; normal or whispered mode & Digital two-speaker mixing with normal and whispered enrollment-target conditions & Yes & Robustness to enrollment-target speech-mode mismatch \\
WHAM!~\cite{WHAM!} & Noisy mixture benchmark & Assigned audio identity & WSJ0-2Mix combined with recorded ambient noise & Yes & Noise-robust extraction and enhancement \\
WHAMR!~\cite{WHAMR!} & Noisy reverberant benchmark & Assigned audio identity & WHAM! sources with simulated reverberation & Yes & Extraction, denoising, and dereverberation \\
SMS-WSJ~\cite{smswsj} & Simulated multichannel benchmark & Spatial geometry & Artificially mixed WSJ speech in randomized simulated rooms and arrays & Yes & Spatial extraction, beamforming, and ASR \\
VoiceBank-DEMAND~\cite{voicedemand} & Adjacent enhancement benchmark & Clean speech; noise & Clean VoiceBank speech digitally mixed with DEMAND noises & Yes & Enhancement evaluation for unified TSE systems \\
FSD50K~\cite{fsd50k} & Sound-event resource & Noise; non-speech events & Independently recorded and labeled sound events & Source known & Interference and noise augmentation \\
\midrule
\rowcolor{gray!20}\multicolumn{6}{@{}c}{\rule{0pt}{2.6ex}\small\textbf{Controlled and natural recorded mixtures}}\\[0.6ex]
LibriCSS~\cite{libricss} & Controlled playback benchmark & Multichannel audio & Conversation-like sequences replayed and recorded in a real room & No exact image & Continuous separation and ASR evaluation \\
AISHELL-4~\cite{Aishell-4} & Natural meeting corpus & Audio; speaker activity & Real Mandarin meetings with natural overlap, noise, and turn-taking & No & Transcription, diarization, adaptation, and robustness \\
CHiME-5~\cite{chime5} & Natural dinner-party corpus & Distributed arrays; binaural audio & Real English conversations recorded in home environments & No exact far-field image & Speaker-dependent enhancement, ASR, and robustness \\
DiPCo~\cite{dipco} & Natural dinner-party corpus & Close-talk and array audio & Real four-speaker conversations recorded with five arrays & No exact far-field image & Meeting transcription and real-mixture evaluation \\
REAL-T~\cite{li2025realt} & Conversation-oriented TSE benchmark & Enrollment speech; transcript & Filtered natural meeting and dinner-party segments paired with target enrollments & No isolated target image & Mandarin and English TSE evaluation using target transcripts \\
LOCATA~\cite{locata} & Controlled spatial corpus & Source trajectories & Static and moving sources recorded with several arrays & Not for TSE & Localization and spatial-cue evaluation \\
\bottomrule
\end{tabularx}
\end{table*}

\begin{table*}[!t]
\caption{Representative cue-bearing multimodal resources used to construct or evaluate visual-, textual-, and neural-conditioned TSE systems. Source clips marked as known usually become exact waveform references only after downstream mixture generation.}
\label{tab:cue_dataset}
\centering
\footnotesize
\setlength{\tabcolsep}{2pt}
\renewcommand{\arraystretch}{1.04}
\begin{tabularx}{\textwidth}{@{}
>{\raggedright\arraybackslash}p{0.15\textwidth}
>{\raggedright\arraybackslash}p{0.145\textwidth}
>{\raggedright\arraybackslash}p{0.12\textwidth}
>{\raggedright\arraybackslash}p{0.255\textwidth}
>{\raggedright\arraybackslash}p{0.105\textwidth}
>{\raggedright\arraybackslash}X@{}}
\toprule
Resource & Primary role & Cue or modality & Mixture or recording construction & Reference status & Typical use in TSE research \\
\midrule
\rowcolor{gray!20}\multicolumn{6}{@{}c}{\rule{0pt}{2.6ex}\small\textbf{Visual and audio-visual resources}}\\[0.6ex]
GRID~\cite{cooke2006audio} & Controlled audio-visual corpus & Face; lip motion & Fixed-format sentences recorded from individual speakers & Source known & Lip-guided extraction and synchronization \\
CUAVE~\cite{CUAVEAN} & Controlled audio-visual corpus & Face; lip motion & Individual and multi-speaker recordings with frontal and profile views & Source-dependent & Early face- and lip-conditioned modeling \\
TCD-TIMIT~\cite{TCD-TIMIT} & Controlled audio-visual corpus & Face; lip motion & Synchronized individual continuous-speech recordings & Source known & Controlled lip-guided extraction \\
OuluVS2~\cite{OuluVS2} & Multi-view audio-visual corpus & Multi-view lip motion & Controlled utterances captured from several camera views & Source known & View-robust visual speech representation \\
LRS2/LRS3~\cite{LRS2,afouras2018lrs3} & In-the-wild audio-visual corpora & Face; lips; transcript & Television and TED or TEDx clips; mixtures generated downstream & Source known & Large-scale lip-guided extraction and synchronization \\
VoxCeleb2~\cite{Voxceleb2} & In-the-wild identity corpus & Face; voice identity & Speaker-labeled online video clips; mixtures generated downstream & Source known & Face identity, enrollment, and mixture generation \\
AVSpeech~\cite{AVSpeech} & In-the-wild audio-visual corpus & Visible speaker & Single-speaker segments collected from online videos & Source known & Large-scale audio-visual mixture generation \\
AVA-ActiveSpeaker~\cite{AVA-ActiveSpeaker} & Active-speaker benchmark & Face track; speaking status & Movie clips with speaking and non-speaking face labels & No clean target & Activity detection and visual-cue pretraining \\
\midrule

\rowcolor{gray!20}\multicolumn{6}{@{}c}{\rule{0pt}{2.6ex}\small\textbf{Textual or semantic resources}}\\[0.6ex]
TextrolSpeech~\cite{ji2024textrolspeech} & Style-annotated source corpus & Natural-language style & Speech paired with rich speaking-style descriptions & Source known & Text encoder and attribute supervision \\
TextrolMix~\cite{huo2025beyond} & Text-conditioned benchmark & Style descriptions & Simulated mixtures paired with target descriptions & Yes & Style-guided target selection \\
PromptTSE~\cite{seki2025language} & Text-conditioned benchmark & Para- and non-linguistic prompts & Prompt-annotated target speech and generated mixtures & Yes & Language-query extraction and prompt generalization \\
SS-Libri~\cite{zhu2025ssdq} & Semantic-spatial benchmark & Text description; region query & Spatialized speech mixtures paired with semantic and spatial queries & Yes & Joint cue disambiguation and fusion evaluation \\
\midrule
\rowcolor{gray!20}\multicolumn{6}{@{}c}{\rule{0pt}{2.6ex}\small\textbf{Neural and attention resources}}\\[0.6ex]
KU Leuven~\cite{das2019auditory} & Neural attention dataset & EEG; attention & Two known speech streams presented during EEG recording & Known stimuli & Neural cue encoding and attended-speaker extraction \\
Cocktail Party~\cite{broderick2018electrophysiological} & Neural speech-tracking dataset & EEG; semantic processing & Natural narrative speech presented during EEG recording & Known stimulus & Neural speech tracking; downstream TSE mixture construction \\
BISS-EEG~\cite{fuglsang2020effects} & Neural attention dataset & EEG; attention & Competing speech presented to normal-hearing and hearing-impaired listeners & Known stimuli & Listener-dependent extraction and hearing-loss analysis \\
BISS-iEEG~\cite{ieeg_data} & Intracranial neural dataset & iEEG; attention & Competing speech presented to clinical participants & Known stimuli & Attention decoding with invasive recordings \\
AVED and MM-AAD~\cite{MSFNet,fan2025seeing} & Multimodal neural datasets & EEG; visual & Audio-visual attention experiments with synchronized neural recordings & Known stimuli & Audio-visual-neural decoding and extraction \\
\bottomrule
\end{tabularx}
\end{table*}

%% file: sections/06_evaluation_metrics.tex
\section{Evaluation Metrics}
\label{sec:evaluation_metrics}

TSE evaluation involves target selection, signal reconstruction, perceptual quality, and application-specific requirements. Since no single metric captures all aspects and reference availability limits valid measures, comprehensive evaluation requires complementary metrics and clear reporting of evaluation conditions.

\subsection{Reference-Based Signal Reconstruction}
When an isolated target signal is available, TSE performance can be evaluated through signal-level comparison with the reference waveform. SDR and its improvement over the mixture, SDRi, are widely reported, although the exact definition and implementation should be specified due to differences in decomposition and alignment conventions. SI-SDR provides a standardized scale-invariant alternative~\cite{le2019sdr}, while SI-SNR and their improvement variants (SI-SDRi/SI-SNRi) are also commonly used to measure reconstruction quality relative to the input mixture.

These metrics quantify waveform fidelity but do not capture all failure modes. Scale-invariant measures are insensitive to overall gain errors, and utterance-level averages may hide short target swaps, leakage, or over-suppression. For continuous or sparsely overlapping speech, evaluation should distinguish target-active regions from inactions to better characterize distortion and residual interference. Therefore, studies should report metric definitions, implementation details, temporal alignment, mixture baselines, and segment selection criteria.

\subsection{Perceptual Quality and Intelligibility}

Signal reconstruction metrics do not always reflect perceived quality or intelligibility. Reference-based measures such as PESQ, STOI, and ESTOI evaluate perceptual quality or intelligibility through comparisons with clean references~\cite{rix2001perceptual,taal2010short,jensen2016estoi}. However, their interpretation depends on the reference signal, bandwidth, and distortion assumptions, and they should not be considered interchangeable.

Human listening tests provide direct assessment of perceived quality, intelligibility, residual interference, and target correctness. Mean opinion scores should be reported with a clear evaluation protocol, listener population, playback condition, and rating criteria~\cite{itu1996p800}. When references are unavailable, non-intrusive learned metrics such as DNSMOS and NISQA estimate perceptual quality directly from the output signal~\cite{reddy2021dnsmos,mittag2021nisqa}. Nevertheless, these estimators may inherit biases from their training domains and should complement rather than replace subjective evaluation.

For speech recognition applications, word error rate (WER) and character error rate (CER) provide task-oriented measures of linguistic preservation. They are particularly useful for real recordings without isolated target references, but depend on the selected recognizer, language model, and decoding protocol. Therefore, ASR evaluation should use fixed recognition settings and include mixture-level baselines for meaningful comparison.

\subsection{Target-Speaker Correctness}

Correct target selection is a defining property of TSE beyond generic speech enhancement. A system may generate clean and intelligible speech while extracting an interfering speaker, especially when target cues are ambiguous or unreliable. Although SI-SDR penalizes complete target swaps, dataset-level averages may obscure rare but critical confusion cases. Explicit target-confusion analysis therefore compares the extracted signal with the target and competing speakers~\cite{zhao2022target}.

Speaker-embedding similarity, verification scores, equal error rate, and target-speaker classification accuracy provide complementary measures of identity preservation and speaker discrimination~\cite{rao2019tse_sv}. Confusion evaluation should report whether the output is more similar to the target than to each interferer, rather than relying only on average similarities. Since these measures depend on the speaker encoder and may be affected by channel or noise conditions, the encoder and evaluation protocol should be fixed and reported. For textual or semantic cues, prompt-output consistency reflects whether the requested information is satisfied, but cannot replace signal- and speaker-level validation when target references are available.

\subsection{Auxiliary-Cue Robustness and Real Recordings}

TSE systems should be evaluated under both ideal and degraded auxiliary cues. Audio enrollment can vary in duration, noise, reverberation, and speaker mismatch; visual cues can suffer from occlusion, low resolution, frame loss, and audio-visual misalignment; spatial cues are affected by localization errors, source motion, reverberation, and array mismatch; textual cues require robustness to paraphrases, ambiguity, and incomplete descriptions; and neural cues should consider subject, session, and alignment variability. In all cases, evaluation should focus on the final extraction quality and target correctness rather than only the accuracy of an intermediate cue estimator.

Real-world recordings often lack isolated target references, making conventional waveform metrics inapplicable. Evaluation should therefore combine available signals, including transcripts, speaker annotations, non-intrusive quality metrics, listening tests, and downstream task performance. For example, ASR evaluates content preservation, speaker verification or diarization evaluates target attribution, and listening tests assess residual interference and artifacts. Results on real recordings should be reported separately from simulated-mixture SI-SDR evaluations, as they reflect different objectives and ground-truth assumptions.

\subsection{Efficiency and Reporting Practice}

Streaming and on-device systems require both causal signal processing and sufficiently fast execution. Algorithmic latency, look-ahead, frame or chunk size, and cue-acquisition delay should be reported separately from computational throughput. Real-time factor indicates whether processing is faster than the signal duration on a specified platform, while parameter count, multiply-accumulate operations or FLOPs, peak memory, and model size describe different aspects of resource use. Reporting RTF without the hardware, numerical precision, batch size, and implementation is insufficient for reproducible comparison. SpeakerBeam-SS illustrates the need to evaluate extraction quality and computational cost jointly~\cite{speakerbeamss}.

Overall, TSE results are most informative when they include at least one reference-based reconstruction measure, one perceptual or intelligibility measure, one target-correctness measure, and, where relevant, a downstream and efficiency measure. Scores should be stratified by factors such as overlap ratio, input signal-to-interference ratio, speaker similarity, noise and reverberation, and cue quality. In addition to the mean, reporting failure rates or distributional statistics helps reveal target confusion and other rare but consequential errors. Table~\ref{tab:libri2mix_independent_reports} catalogues independently reported Libri2Mix results to show the breadth of metrics used in the literature. Because the cited studies differ in data construction, training resources, model scale, and evaluation implementation, those values should not be interpreted as a controlled ranking. Table~\ref{tab:libri2mix_discriminative_generative} separately presents the LauraTSE comparison under a common evaluation protocol on Libri2Mix Clean~\cite{lauratse}. Its caption records the evaluation models and identifies published, adjacent, and preprint results.

\input{table_DG}

%% file: table_DG.tex
\begin{table*}[!t]
\centering
\caption{Catalogue of independently reported results on Libri2Mix. Values are transcribed from the cited studies and are listed to summarize the range of reported evaluations, not to establish a controlled ranking. Papers differ in training data, mixture recipes, clean or noisy test subsets, sampling rates, model scale, and metric implementations. The SI-SDR/SI-SNR column retains the signal-level measure named by each source. D, G, and H denote discriminative, generative, and hybrid discriminative-generative systems. $\dagger$ marks an adjacent target-sound system and $\ddagger$ marks a preprint-only system. NR indicates that the source did not report the metric.}
\label{tab:libri2mix_independent_reports}
\footnotesize
\renewcommand{\arraystretch}{1.05}
\setlength{\tabcolsep}{2.5pt}
\begin{tabular*}{\textwidth}{@{\extracolsep{\fill}}llcccccc@{}}
\toprule
Method & Cat. & PESQ $\uparrow$ & ESTOI $\uparrow$ & \makecell{SI-SDR/\\SI-SNR $\uparrow$} & DNSMOS $\uparrow$ & WER $\downarrow$ & SIM $\uparrow$ \\
\midrule
\rowcolor{gray!20}\multicolumn{8}{@{}c}{\rule{0pt}{2.6ex}\small\textbf{Libri2Mix Noisy}}\\[0.6ex]
TD-SpeakerBeam~\cite{delcroix2020improving} & D & 1.66 & 0.70 & 9.21 & 3.14 & 0.21 & 0.93 \\
X-TF-GridNet~\cite{XTFGRIDNET} & D & 1.72 & 0.72 & 9.85 & 3.42 & 0.19 & 0.93 \\
SSL-MHFA~\cite{sslmhfa} & D & 1.76 & 0.74 & 10.60 & 3.22 & 0.17 & 0.94 \\
USEF-TSE~\cite{usef} & D & 1.82 & 0.72 & 10.17 & 3.48 & 0.17 & 0.94 \\
AD-FlowTSE~\cite{hsieh2025adaptive} & G & 2.15 & 0.81 & 12.69 & 3.48 & NR & 0.87 \\
\midrule
DDTSE~\cite{ddtse} & H & 1.60 & 0.71 & 7.60 & 3.74 & NR & 0.71 \\
Diff-TSE~\cite{difftse} & G & 1.56 & 0.64 & 8.35 & 3.53 & 0.25 & 0.89 \\
FlowTSE~\cite{flowtse} & G & 1.86 & 0.75 & NR & 3.82 & NR & 0.83 \\
SoloAudio$^{\dagger}$~\cite{wang2025soloaudio} & G & 1.56 & 0.63 & 2.64 & 3.71 & 0.35 & 0.93 \\
SoloSpeech$^{\ddagger}$~\cite{solospeech} & G & 1.89 & 0.78 & 11.12 & 3.76 & 0.15 & NR \\
\midrule
\rowcolor{gray!20}\multicolumn{8}{@{}c}{\rule{0pt}{2.6ex}\small\textbf{Libri2Mix Clean}}\\[0.6ex]
AD-FlowTSE~\cite{hsieh2025adaptive} & G & 2.89 & 0.90 & 17.49 & 3.59 & NR & 0.95 \\
DDTSE~\cite{ddtse} & H & 1.79 & 0.78 & NR & 3.79 & NR & 0.73 \\
Diff-TSE~\cite{difftse} & G & 3.08 & 0.80 & 11.28 & NR & NR & NR \\
FlowTSE~\cite{flowtse} & G & 2.58 & 0.84 & NR & 3.79 & NR & 0.90 \\
SR-SSL~\cite{srssl} & G & 2.99 & NR & 16.00 & NR & NR & NR \\
\bottomrule
\end{tabular*}
\end{table*}

\begin{table*}[!t]
\centering
\caption{Perceptual, linguistic, and speaker-consistency results on Libri2Mix Clean reported under the LauraTSE evaluation protocol~\cite{lauratse}. D, G, and D-G denote discriminative, generative, and discriminative-generative systems. $\dagger$ identifies a unified restoration system rather than a dedicated TSE model, and $\ddagger$ marks results from a preprint extension. LauraTSE evaluates dWER with Whisper-base, SpeechBERT with HuBERT-base representations, and speaker similarity with WavLM-base-plus-SV and the WeSpeaker ResNet-221LM. The LauraTSE paper reports that the AnyEnhance values were taken from the original AnyEnhance study, whereas the remaining published baseline outputs were evaluated in its comparison.}
\label{tab:libri2mix_discriminative_generative}
\footnotesize
\renewcommand{\arraystretch}{1.05}
\setlength{\tabcolsep}{2.5pt}
\begin{tabular*}{\textwidth}{@{\extracolsep{\fill}}lccccccccc@{}}
\toprule
\multirow{2}{*}{Method}
& \multirow{2}{*}{Cat.}
& \multicolumn{3}{c}{DNSMOS $\uparrow$}
& \multirow{2}{*}{NISQA $\uparrow$}
& \multirow{2}{*}{\makecell{SpeechBERT\\$\uparrow$}}
& \multirow{2}{*}{dWER $\downarrow$}
& \multirow{2}{*}{\makecell{WavLM\\Sim. $\uparrow$}}
& \multirow{2}{*}{\makecell{WeSpeaker\\Sim. $\uparrow$}}\\
\cmidrule(lr){3-5}
& & SIG & BAK & OVRL & & & & & \\
\midrule
Mixture
& N/A
& 3.383
& 3.098
& 2.653
& 2.453
& 0.572
& 0.792
& 0.847
& 0.759\\
\midrule
SpEx+~\cite{spex+}
& D
& 3.472
& 4.027
& 3.186
& 3.349
& 0.878
& 0.148
& 0.973
& 0.935\\
WeSep~\cite{wesep}
& D
& 3.486
& 3.838
& 3.118
& 3.892
& 0.895
& 0.123
& 0.980
& 0.945\\
USEF-TSE~\cite{usef}
& D
& 3.555
& 4.051
& 3.272
& 4.319
& 0.935
& 0.075
& 0.988
& 0.968\\
\midrule
TSELM-L~\cite{tselm}
& G
& 3.489
& 4.041
& 3.212
& 3.961
& 0.793
& 0.297
& 0.887
& 0.627\\
AnyEnhance$^{\dagger}$~\cite{anyenhance}
& G
& 3.638
& 4.066
& 3.353
& 4.277
& 0.735
& NR
& 0.914
& NR\\
LauraTSE~\cite{lauratse}
& G
& 3.609
& 4.084
& 3.336
& 4.333
& 0.908
& 0.159
& 0.974
& 0.876\\
LauraTSE-Streaming~\cite{lauratse}
& G
& 3.596
& 4.061
& 3.314
& 4.275
& 0.897
& 0.169
& 0.973
& 0.873\\
\midrule
\rowcolor{gray!20}\multicolumn{10}{@{}c}{\rule{0pt}{2.6ex}\small\textbf{Discriminative-generative preprint extension$^{\ddagger}$}}\\[0.6ex]
USEF-Laura-TSE-S (G, no SI-SDR)~\cite{useflauratse}
& D-G$^{\ddagger}$ & 3.592 & 4.061 & 3.313 & 4.453 & 0.925 & 0.120 & 0.978 & 0.895\\
USEF-Laura-TSE-S (D)~\cite{useflauratse}
& D-G$^{\ddagger}$ & 3.422 & 3.661 & 2.979 & 3.172 & 0.884 & 0.113 & 0.977 & 0.937\\
USEF-Laura-TSE-S (G)~\cite{useflauratse}
& D-G$^{\ddagger}$ & 3.603 & 4.080 & 3.329 & 4.416 & 0.915 & 0.154 & 0.975 & 0.880\\
USEF-Laura-TSE-L (D)~\cite{useflauratse}
& D-G$^{\ddagger}$ & 3.528 & 3.955 & 3.202 & 3.648 & 0.933 & 0.076 & 0.987 & 0.950\\
USEF-Laura-TSE-L (G)~\cite{useflauratse}
& D-G$^{\ddagger}$ & 3.592 & 4.075 & 3.319 & 4.450 & 0.934 & 0.117 & 0.982 & 0.902\\
\bottomrule
\end{tabular*}
\end{table*}

%% file: sections/07_challenges_open_problems.tex
\section{Challenges and Open Problems}
\label{sec:challenge}

The preceding sections show that TSE performance depends on more than the capacity of the extraction network. A practical system must determine whether the auxiliary cue is available, whether it specifies a unique source in the current scene, and whether the resulting output remains reliable as speakers, sensors, and acoustic conditions change.

\subsection{Benchmark Realism, Data Coverage, and Comparability}

Most supervised TSE benchmarks retain isolated target signals by digitally combining independently recorded utterances. This enables controlled training and reference-based evaluation, while not reproducing all properties of natural conversation, including irregular overlap, turn taking, correlated noise, source movement, and interaction with the room. Simulated spatial benchmarks additionally depend on the room, array, and source-position distributions selected by the data-generation recipe~\cite{librimix,smswsj}. In contrast, recorded meeting corpora capture realistic conversations and propagation conditions but generally do not provide the exact isolated target image at each far-field microphone~\cite{libricss,Aishell-4,chime5,dipco}. The field therefore faces a persistent trade-off between exact supervision and acoustic realism.

Data scarcity and coverage are also uneven across modalities. Audio-only source corpora can contain thousands of speakers, whereas synchronized visual and especially neural datasets involve fewer participants, sessions, languages, and recording environments. Neural data collection further requires specialized hardware and participant-specific recording procedures~\cite{das2019auditory,fuglsang2020effects,ieeg_data}. Even when studies use the same named speech corpus, differences in enrollment selection, overlap, relative level, noise, reverberation, spatialization, and speaker partitioning can make their results non-equivalent. Consequently, benchmark scale alone does not establish diversity, realism, or comparability.

\subsection{Cue Availability, Reliability, Synchronization, and Conflict}

Every auxiliary cue has a distinct failure process. Enrollment speech may be noisy, reverberant, too short, or recorded through a mismatched channel. Faces and gestures may be occluded, outside the camera view, or incorrectly tracked. Lip motion and neural activity require temporal alignment with the acoustic mixture. Spatial cues depend on localization accuracy, array calibration, and separation between target and interfering directions. Textual descriptions may be incomplete or inconsistent with the speakers present. Neural observations vary across listeners, sessions, and electrode configurations.

Combining modalities does not remove these problems automatically. One cue may be reliable while another is missing, delayed, or points to a different source. Fixed fusion can then amplify an incorrect observation or cause the model to rely primarily on the easiest modality during training. Existing studies have examined reliability-aware audio-visual attention, modality dropout, impaired visual streams, perturbed direction estimates, and missing lip or gesture observations~\cite{sato2021multimodal,korse2024modality,li2024momuse,wang2024study,pan2026beyondlips}. However, these studies use different corruption models and operating conditions. A general unresolved issue is therefore how to estimate cue reliability, align cues with different temporal resolutions, associate each observation with the same physical speaker, and respond when the cues contradict one another.

\subsection{Target Ambiguity, Confusion, and Absence}

An auxiliary cue does not always define a unique target. Speakers may have similar voice characteristics, occupy similar directions, share a textual attribute, or appear simultaneously in the camera view. A relational instruction is meaningful only when the relevant contrast exists, and a spatial region may contain more than one speaker. Neural attention may also be uncertain during attention switches or when competing streams evoke similar responses. In these cases, the separator is asked to make a target assignment from insufficient or ambiguous evidence.

This ambiguity can produce target confusion, in which the output contains the wrong speaker despite sounding like clean speech. Target confusion has been observed for enrollment-conditioned extraction and is not fully characterized by an average reconstruction score~\cite{zhao2022target}. The problem also extends to target-inactive and target-absent intervals. A system must preserve the target when active, suppress interfering speech when the target is silent, and avoid producing a speaker when the requested target is not present~\cite{borsdorf2021universal,pan2022usev}. These requirements become harder in continuous recordings because the target may enter, leave, move, or change speaking activity without an explicit boundary. Rejection and uncertainty estimation are therefore separate requirements from closed-set extraction in which a valid target is assumed.

\subsection{Generalization Across Speakers, Domains and  Devices}

Generalization in multimodal TSE involves several shifts at once. The acoustic mixture may contain unseen speakers, languages, noises, rooms, and microphones, while the target cue may come from a new camera, array geometry, text style, or neural-recording session. A model can generalize acoustically yet fail because its cue encoder or fusion mechanism does not transfer. Conversely, a robust speaker or visual representation does not guarantee that the separator will use it correctly under a new mixture condition. Controlled comparisons have shown that the benefit of auxiliary audio or visual information depends on the dataset, number of speakers, input condition, and cue representation~\cite{elminshawi2024auxiliary}.

The generalization problem is especially pronounced for neural conditioning because EEG characteristics vary across listeners and sessions, and many studies use subject-dependent training or evaluation~\cite{pan2023neuroheed,neurospex,m3anet}. Spatial systems face geometry and calibration mismatch, while textual systems must connect linguistic variation to acoustic attributes across languages and speaking styles. Multimodal systems add another layer of shift because the reliability relationship learned between cues may change after deployment. Performance on unseen acoustic mixtures is therefore only one component of cross-domain robustness.

\subsection{Continuous and Dynamic Interaction}

Many benchmark mixtures contain a fixed set of speakers and are processed as independent utterances. Natural interaction instead includes partial overlap, long single-speaker regions, silence, interruptions, moving sources, visual occlusion, and changes in the intended target. A continuous TSE system must maintain target identity across these events while avoiding leakage during target-inactive periods. It must also decide when stored identity information should be retained and when it should be updated after a user changes attention.

Recent systems have begun to study sparsely overlapping speech, impaired visual streams, moving speakers, and online target changes~\cite{li2024activeextract,pan2022usev,li2024momuse,avase,kienegger2025steering}. These settings expose problems that are largely absent from short fully overlapped mixtures: accumulated identity drift, delayed cue updates, error propagation from previously extracted speech, and instability when speakers cross or alternate. Causal processing further restricts future context, while visual, spatial, or neural cue estimators may introduce their own delay. Continuous operation is therefore both a target-tracking problem and a signal-reconstruction problem.

\subsection{Evaluation Validity and Application Dependence}

Sec.~\ref{sec:evaluation_metrics} shows that signal fidelity, perceived quality, intelligibility, speaker identity, and downstream performance measure different properties. A high perceptual-quality score does not establish that the correct speaker was selected, and a high average SI-SDR can hide a small number of severe target-confusion failures. On natural recordings without isolated targets, waveform metrics cannot be computed directly, while non-intrusive quality estimators do not measure target attribution. The relative importance of these outcomes also depends on whether the output is intended for human listening, automatic speech recognition, speaker verification, or another downstream system.

Cross-cue comparison is particularly difficult because cue types often use different datasets and assumptions. Audio enrollment is commonly evaluated on speaker-disjoint simulated mixtures, visual systems require synchronized video, spatial systems use particular array geometries, textual systems define their own prompt inventories, and neural studies use participant-specific recordings. A single numerical ranking across these settings would confound cue utility with dataset difficulty, sensing conditions, backbone capacity, and training data. The open problem is not merely selecting more metrics, but establishing protocols that reveal target-selection failures and permit controlled comparisons without erasing modality-specific requirements.

\subsection{Computational Cost and End-to-End Latency}

Real-time TSE requires more than a causal separator. The complete processing path can include cue acquisition, face or lip tracking, speaker or text encoding, localization, neural-signal preprocessing, multimodal fusion, waveform extraction, and output reconstruction. Audio enrollment embeddings may be computed before extraction, whereas synchronized visual, spatial, and neural cues must often be updated online. Reported separator latency therefore may not represent the delay experienced by the user.

Network design introduces additional trade-offs. Large pretrained encoders increase memory and computation, diffusion and flow systems may require several model evaluations, and codec-based systems add tokenization and waveform decoding. Lightweight causal systems demonstrate that real-time extraction is possible, but performance, algorithmic latency, throughput, memory, and power consumption remain coupled~\cite{uninet,personalized,speakerbeamss}. Comparisons are further complicated when real-time factor is reported on different hardware or without including cue-processing costs. This limits conclusions about readiness for hearing aids, wearable devices, teleconferencing, or other resource-constrained applications.

\subsection{Privacy, Consent, and Responsible Use}

The target cues reviewed in this paper can reveal sensitive information. Enrollment speech and facial observations encode identity, spatial cues reveal location, textual prompts may expose personal attributes or conversational context, and neural signals are closely tied to an individual listener and experimental session. Collecting several modalities together increases the amount of information that can be linked to one person. These risks concern not only model training but also storage, transmission, reuse, and access to embeddings or intermediate sensor data.

TSE can support communication and hearing assistance, but the same ability to isolate a selected person can also be used without that person's knowledge. Consent may be especially difficult in meetings or public spaces containing several speakers, only one of whom operates the extraction system. Dataset imbalance can additionally produce uneven performance across languages, accents, demographic groups, hearing conditions, or neural-recording populations. Conventional extraction metrics do not measure these risks. Privacy, consent, access control, demographic coverage, and potential misuse must therefore be treated as deployment requirements rather than inferred from signal quality alone.

These challenges are interdependent. More modalities can reduce target ambiguity but increase synchronization, computation, and privacy burdens; more realistic recordings improve ecological validity but reduce the availability of isolated references; and more powerful generative or pretrained models can improve representation capacity while complicating fidelity and deployment assessment. Section~\ref{sec:future_direction} discusses research directions that respond to these tensions.

%% file: sections/08_future_directions.tex
\section{Future Directions}
\label{sec:future_direction}

The challenges in Sec.~\ref{sec:challenge} suggest that future progress will depend on treating target specification, extraction, evaluation, and deployment as a connected problem. The directions below extend trends already visible in the reviewed literature while identifying capabilities that remain incomplete.

\subsection{Adaptive and Reliability-Aware Cue Fusion}

Future multimodal TSE systems should estimate not only a representation for each cue but also its reliability for the current time interval. Such a system could down-weight a noisy enrollment utterance, bridge a short visual occlusion, reject an inaccurate direction estimate, or defer to a second cue when two observations disagree. Initial studies have explored reliability-aware attention, modality dropout, explicit cue-validity classification, and complementary lip and gesture streams~\cite{sato2021multimodal,korse2024modality,eisenberg2026spectralspatial,pan2026beyondlips}. A broader framework should generalize these ideas beyond a fixed pair of modalities.

One promising formulation is to maintain modality-specific encoders and a shared fusion module that predicts calibrated cue confidence alongside the target estimate. Training should include missing, delayed, corrupted, and contradictory cues, rather than only complete observations. Because audio identity, lip motion, spatial position, text, and neural attention operate at different temporal resolutions, confidence should be time varying and tied to explicit cross-modal association. The model should also be able to operate with any available subset of cues without requiring retraining for every combination. Evaluation would then measure both extraction quality and whether estimated confidence predicts target-selection failure.

\subsection{Instruction-Driven and Ambiguity-Aware Extraction}

Natural-language interaction can provide a common interface to heterogeneous target cues. An instruction may refer to identity, position, speaking style, content, a relation between speakers, or dialogue context. Existing text-guided systems demonstrate extraction from free-form descriptions, para-linguistic attributes, inter-speaker relations, and conversational history~\cite{hao2025typing,seki2025language,dai2025inter,kim2025contextual}. Large language models and multimodal language models can serve as instruction parsers or context encoders, but they do not replace evidence that identifies the target in the observed scene. The next step is therefore not merely using a larger text encoder, but grounding each instruction in available acoustic, visual, spatial, or contextual evidence.

An instruction-driven system could parse a request into several constraints, such as speaker identity, relative location, and current speaking activity, and then connect each constraint to the relevant sensor or cue encoder. When several speakers satisfy the request, the target is absent, or the instruction conflicts with the observations, the system should expose uncertainty, abstain, or request clarification instead of producing an arbitrary speaker. Joint target-presence detection and extraction provides an early example of separating rejection from waveform reconstruction~\cite{borsdorf2021universal,zhang2023tsejoint}. Interactive TSE should therefore be evaluated on ambiguous, unsatisfiable, paraphrased, and multi-turn instructions, in addition to prompts that uniquely identify a target.

\subsection{Continuous Target Tracking in Dynamic Scenes}

Future TSE should move from independent utterance processing toward persistent target tracking. The system state may connect identity, activity, and location across silence, partial overlap, visual interruption, speaker movement, and changes in user attention. Memory-based visual conditioning, autoregressive acoustic guidance, weakly guided spatial tracking, and combined EEG and eye-tracking cues provide early components for this setting~\cite{li2024momuse,avase,kienegger2025steering,wang2026eegeye}.

A continuous system should distinguish between temporary cue loss and a genuine change of target. This requires an explicit policy for retaining, updating, or resetting the target representation. State updates derived from previously extracted speech should be confidence gated to prevent extraction errors from contaminating future estimates. Dynamic evaluation should include target entry and exit, speaker crossing, attention switching, target-inactive intervals, and long recordings in which small frame-level errors may accumulate. Such protocols would connect TSE more closely with active-speaker detection, target-speaker activity detection, localization, and diarization without requiring these tasks to share a single output representation.

\subsection{Pretrained and Unified Models}

Pretrained speech and audio models supply stronger semantic, speaker, and acoustic representations, while unified generative systems transfer knowledge across enhancement, separation, extraction, editing, and synthesis~\cite{nemo,speechx,uniaudio,anyenhance,llaseg1}. Their future value for TSE  depend on whether this transfer preserves the properties that distinguish extraction from general restoration: correct target assignment, suppression during target absence, fidelity to the observed speaker, and controllable latency.

A practical direction is to share representations from large pretrained encoders while retaining lightweight TSE-specific modules for cue grounding and target reconstruction. This design can support speech enhancement, separation, and TSE within a common backbone without sacrificing task-specific conditioning and output constraints.
Parameter-efficient adaptation, distillation, caching, and conditional activation can reduce the cost of repeatedly running large encoders. For generative models, perceptual restoration should be balanced against exact recovery through mixture or speaker consistency and uncertainty estimation. Unified models also require task-specific evaluation, as average performance across speech transformations can obscure failures in TSE.

\subsection{Realistic Multimodal Benchmarks and Evaluation}

Future benchmarks should connect controlled simulation with natural conversation instead of treating them as competing alternatives. Synthetic mixtures remain useful for exact target references and controlled stress tests, while recorded conversations capture spontaneous overlap, source movement, room effects, and sensor failures. Conversation-oriented resources such as REAL-T provide a step toward evaluating TSE under more realistic interaction patterns~\cite{li2025realt}. A stronger benchmark design could pair or align distant recordings with close-talking references, transcripts, speaker activity, face tracks, locations, prompts, or attention labels whenever these annotations are feasible.

Benchmark protocols should standardize speaker and session partitions, acoustic conditions, cue corruption, and target presence. Results can be stratified by overlap, speaker similarity, acoustic mismatch, cue quality, and target activity. Evaluation should go beyond signal fidelity to measure target confusion, perceptual quality and intelligibility, downstream performance, latency, and user experience when relevant. Controlled cross-cue comparisons require a shared acoustic backbone or matched parameter and training-data budgets to isolate the contribution of auxiliary cues~\cite{elminshawi2024auxiliary}. Public generation and evaluation code can further improve reproducibility.

\subsection{Causal and Resource-Efficient End-to-End Systems}

Deployment-oriented research should optimize the complete sensing and extraction pipeline rather than the separator alone. This includes cue acquisition, face tracking or localization, cue encoding, fusion, waveform estimation, and reconstruction. Causal TSE models have already demonstrated low-latency recurrent, perceptual, and state-space designs~\cite{uninet,personalized,speakerbeamss}. Future systems should extend this efficiency to multimodal frontends and dynamic cue selection.

Several complementary strategies are relevant: caching slowly changing identity or text embeddings, updating high-rate cues only when necessary, using confidence to skip unavailable modalities, compressing large encoders, and adapting computation to scene difficulty. Hardware-aware training and evaluation should report end-to-end algorithmic latency, cue delay, real-time factor, memory, numerical precision, and power on the intended platform. For hearing-assistive or wearable applications, the appropriate optimization target may be a constrained trade-off among target correctness, perceptual quality, battery use, and delay rather than the highest offline reconstruction score.

\subsection{Trustworthy and Privacy-Preserving Deployment}

Trustworthy TSE requires technical and procedural safeguards throughout the system lifecycle. Sensitive cue processing should be minimized, and identity, facial, location, textual, or neural representations should not be retained longer than required by the application. On-device processing, protected embeddings, access control, and privacy-preserving training are possible components, but their usefulness must be evaluated together with extraction accuracy and computational cost.

Users should be able to understand which cues are active, change or revoke the selected target, and recognize when the system is uncertain. Systems intended for meetings, assistive listening, or public environments also need consent and recording policies that consider speakers other than the device owner. Evaluation should report performance across languages, accents, demographic groups, hearing conditions, devices, and, for neural systems, participants and sessions. Failure analysis should include incorrect-speaker extraction and unauthorized or unintended target selection, not only signal degradation. These practices would make trustworthy deployment a measurable design objective rather than a general statement about responsible use.

Together, these directions point toward TSE systems that can interpret user intent, combine available evidence, track a target over time, and communicate uncertainty while satisfying application-specific fidelity, latency, and privacy requirements. Progress will require coordinated advances in cue modeling, benchmark design, evaluation, and system integration rather than improvement of the extraction backbone alone.

%% file: sections/09_conclusion.tex
\section{Conclusion}
This review examined  TSE from the perspective of auxiliary target cues drawn from multiple modalities. We organized existing methods according to audio, visual, spatial, textual or semantic, and neural cues, and described the information that each cue provides for target identification, its incorporation into extraction systems, and its practical requirements and limitations. We also reviewed the evolution of TSE architectures from discriminative estimators to variational, diffusion, flow, codec, pretrained, and unified models. In addition, we distinguished source corpora from constructed and recorded TSE benchmarks and summarized evaluation measures for reconstruction quality, intelligibility, target-speaker correctness, cue robustness, and computational efficiency.
The reviewed literature indicates that no auxiliary cue or model paradigm is universally suitable. Their usefulness depends on the application, available sensors, cue reliability, acoustic conditions, latency requirements, and privacy constraints. Important unresolved problems include incomplete or conflicting cues, target ambiguity and absence, generalization across speakers and recording conditions, continuous target tracking, limited real-mixture evaluation, and inconsistent reporting protocols. Future progress therefore requires reliability-aware cue fusion, grounded and ambiguity-aware instructions, realistic multimodal benchmarks, efficient causal systems, and explicit safeguards for privacy and trustworthy use. Addressing these issues can support TSE to identify the intended speaker more reliably while meeting the fidelity, efficiency, and deployment requirements of real-world speech applications.

%% file: table_method.tex
\onecolumn
\begin{landscape}
\footnotesize
\setlength{\LTleft}{0pt}
\setlength{\LTright}{0pt}
\setlength{\LTcapwidth}{\linewidth}
\setlength{\tabcolsep}{6pt}
\renewcommand{\arraystretch}{1.08}
\begin{xltabular}{\linewidth}{@{}l c l l >{\raggedright\arraybackslash}X@{}}
\caption{Taxonomy of representative systems according to the auxiliary information used for target selection. The table intentionally reports cue roles and system scope rather than compact implementation fields whose definitions differ across papers. Detailed architectures and objectives are discussed in Sec.~\ref{sec:type_cues} and~\ref{sec:evolution}. $\dagger$ identifies adjacent or unified systems included for a transferable conditioning or modeling mechanism.}
\label{tab:sota}\\
\toprule
System & Year & Auxiliary target cue & Evidence used to select the target & Scope \\
\midrule
\endfirsthead
\toprule
System & Year & Auxiliary target cue & Evidence used to select the target & Scope \\
\midrule
\endhead
\midrule
\multicolumn{5}{r}{\textit{Continued on next page}}\\
\endfoot
\bottomrule
\endlastfoot

\rowcolor{gray!20}\multicolumn{5}{@{}c}{\rule{0pt}{2.6ex}\small\textbf{Audio enrollment and reference-conditioned systems}}\\[0.6ex]
Single-channel SpeakerBeam~\cite{delcroix2018speakerbeam} & 2018 & Enrollment utterance & Jointly learned speaker-aware adaptation & Dedicated TSE \\
SpeakerBeam~\cite{speakerbeam} & 2019 & Enrollment utterance & Persistent speaker identity & Dedicated TSE \\
VoiceFilter~\cite{wang2019voicefilter} & 2019 & Enrollment utterance & Speaker-verification embedding & Dedicated TSE \\
Time-domain SpeakerBeam~\cite{delcroix2020improving} & 2020 & Enrollment utterance & Speaker-discriminative waveform conditioning & Dedicated TSE \\
Joint speaker-aware extraction~\cite{ji2020speakeraware} & 2020 & Enrollment utterance & Jointly optimized speaker embedding & Dedicated TSE \\
SpEx~\cite{spex} & 2020 & Enrollment utterance & Jointly learned speaker identity & Dedicated TSE \\
SpEx+~\cite{spex+} & 2020 & Enrollment utterance & Speaker identity in a shared representation space & Dedicated TSE \\
SpEx-CAN~\cite{spexcan} & 2021 & Enrollment utterance & Frame-level reference and mixture interaction & Dedicated TSE \\
Personalized PercepNet~\cite{personalized} & 2021 & Enrollment utterance & Personalized speaker embedding & Causal dedicated TSE \\
GCF-TSE~\cite{gcntse} & 2022 & Enrollment utterance & Gated speaker conditioning & Dedicated TSE \\
Conformer TSE~\cite{sinha2022conformer} & 2022 & Enrollment utterance & Convolution and self-attention speaker conditioning & Dedicated TSE \\
CIENet~\cite{cienet} & 2024 & Enrollment utterance & Contextual interaction between mixture and enrollment & Dedicated TSE \\
SSL-MHFA~\cite{sslmhfa} & 2024 & Enrollment utterance & Pretrained speech and speaker representations & Dedicated TSE \\
Centroid-TSE~\cite{heo2024centroid} & 2024 & Enrollment utterance & Enrollment-derived speaker centroid & Dedicated TSE \\
Hierarchical speaker representation~\cite{he2024hierarchical} & 2024 & Enrollment utterance & Speaker information from multiple network levels & Dedicated TSE \\
Coarse-to-fine TSE~\cite{yang2024coarsetofine} & 2024 & Enrollment utterance & Successive contextual extraction resolutions & Dedicated TSE \\
Unified attractor-based SS and TSE~\cite{chetupalli2024unified} & 2024 & Acoustic, semantic, or visual clue & User-conditioned selection with encoder-decoder attractors & Unified separation and TSE \\
OR-TSE~\cite{zhang2024ortse} & 2024 & Overlap-contaminated enrollment & Target identity under enrollment interference & Dedicated TSE \\
SkiM-UniCATS~\cite{skim} & 2024 & Enrollment utterance & Reference-conditioned discrete speech modeling & Dedicated TSE \\
TSE-PENE~\cite{tsepene} & 2025 & Positive and negative noisy enrollments & Target evidence obtained by contrasting enrollment segments & Dedicated TSE \\
USEF-TSE~\cite{usef} & 2025 & Enrollment utterance & Reference-to-mixture cross-attention and modulation & Dedicated TSE \\
MLTSE~\cite{mltse} & 2025 & Enrollment utterance & Multi-level target-speaker information & Dedicated TSE \\
DSINet~\cite{hao2025dsinet} & 2025 & Enrollment utterance & Dynamic speaker-information fusion & Real-time dedicated TSE \\
DCF-Net~\cite{xue2025dcf} & 2025 & Enrollment utterance & Contextual mixture-enrollment interaction & Dedicated TSE \\
TargetVoice~\cite{pallala2025targetvoice} & 2025 & Enrollment utterance & Compact low-latency speaker conditioning & Real-time dedicated TSE \\
MTSE~\cite{serre2025mtse} & 2025 & Multiple enrollment utterances & A selected set of enrolled speakers & Multi-target speaker extraction \\
LExt~\cite{shen2025lext} & 2025 & Prepended enrollment waveform & Artificial-onset target prompting & Dedicated TSE \\
LauraTSE~\cite{lauratse} & 2025 & Enrollment utterance & Reference-conditioned codec representation & Dedicated TSE \\
SpeechX$^{\dagger}$~\cite{speechx} & 2024 & Enrollment utterance and task instruction & Target identity within a unified speech transformation model & Unified speech processing \\
UniAudio$^{\dagger}$~\cite{uniaudio} & 2024 & Enrollment utterance and task token & Target identity within a unified audio generator & Unified audio generation \\
Metis$^{\dagger}$~\cite{wang2025metis} & 2025 & Enrollment utterance and task condition & Target identity within a pretrained speech generator & Speech foundation model \\
LLaSE-G1$^{\dagger}$~\cite{llaseg1} & 2025 & Enrollment utterance and instruction & Reference-conditioned speech restoration & Unified speech restoration \\
AnyEnhance$^{\dagger}$~\cite{anyenhance} & 2025 & Enrollment utterance and prompt & Reference-conditioned enhancement or extraction & Unified voice restoration \\
TSELM~\cite{tselm} & 2026 & Enrollment utterance & Reference-conditioned discrete target tokens & Dedicated TSE \\

\midrule
\rowcolor{gray!20}\multicolumn{5}{@{}c}{\rule{0pt}{2.6ex}\small\textbf{Visual and audio-visual systems}}\\[0.6ex]
AV-ConvTasNet~\cite{wu2019time} & 2019 & Synchronized lip motion & Time-varying articulation & Dedicated AV-TSE \\
Multimodal SpeakerBeam~\cite{ochiai2019multimodal} & 2019 & Enrollment utterance and face video & Voice identity and visible articulation & Multimodal TSE \\
FaceFilter~\cite{chung2020facefilter} & 2020 & Still face image & Persistent facial identity & Dedicated AV-TSE \\
MuSE~\cite{pan2021muse} & 2021 & Synchronized lip motion & Articulation and self-enrolled voice identity & Dedicated AV-TSE \\
Multimodal attention fusion~\cite{sato2021multimodal} & 2021 & Enrollment utterance and face video & Reliability-weighted voice and visual evidence & Multimodal TSE \\
Reentry~\cite{pan2021reentry} & 2022 & Synchronized lip motion & Pretrained speech-lip synchronization & Dedicated AV-TSE \\
USEV~\cite{pan2022usev} & 2022 & Synchronized lip motion & Articulation during sparse and overlapping speech & Dedicated AV-TSE \\
SEG~\cite{pan2022seg} & 2022 & Co-speech gesture & Time-varying behavioral correlation with speech & Gesture-guided TSE \\
VCSE~\cite{li2022vcse} & 2022 & Lip motion and self-enrolled context & Articulation and phonetic context & Dedicated AV-TSE \\
LiMuSE~\cite{liu2023limuse} & 2023 & Face video and voiceprint & Visual activity and speaker identity & Multimodal TSE \\
AV-SepFormer~\cite{Lin2023sepformer} & 2023 & Synchronized lip motion & Cross-modal articulatory evidence & Dedicated AV-TSE \\
SAV-GridNet~\cite{pan2023scenario} & 2023 & Synchronized lip motion & Articulation under scenario-dependent conditions & Dedicated AV-TSE \\
ImagineNet~\cite{pan2023imaginenet} & 2023 & Synchronized lip motion & Audio-visual correspondence & Dedicated AV-TSE \\
DAVSE~\cite{li2023rethinking} & 2023 & Face identity and lip synchronization & Separated identity and synchronization evidence & Dedicated AV-TSE \\
PIAVE~\cite{liu2023piave} & 2023 & Face video under pose variation & Original and pose-normalized facial evidence & Dedicated AV-TSE \\
Dual-path cross-modal TSE~\cite{xu2023dual} & 2023 & Face features & Dual-path audio-visual cross-attention & Dedicated AV-TSE \\
AV-HuMAR~\cite{wu2024avmar} & 2024 & Synchronized lip motion & Pretrained visual-speech representation & Dedicated AV-TSE \\
AVSepChain~\cite{mu2024speechchain} & 2024 & Synchronized lip motion & Reciprocal speech-perception and reconstruction cues & Dedicated AV-TSE \\
ActiveExtract~\cite{li2024activeextract} & 2024 & Audio-visual active-speaker features & Speaking activity and synchronization & Dedicated AV-TSE \\
UniNet~\cite{wu2024unified} & 2024 & Enrollment utterance and lip motion & Voice identity and articulation & Multimodal TSE \\
MoMuSE~\cite{li2024momuse} & 2025 & Face video with temporal memory & Persistent identity across impaired frames & Online AV-TSE \\
$C^{2}$AV-TSE~\cite{wu2025c} & 2025 & Lip motion, context, and confidence & Contextual visual evidence and estimated reliability & Dedicated AV-TSE \\
AV-CrossNet$^{\dagger}$~\cite{kalkhorani2025avcrossnet} & 2025 & Synchronized lip motion & Articulatory conditioning for several AV speech tasks & Multi-task AV speech processing \\
SEANet~\cite{tao2025audio} & 2025 & Synchronized lip motion & Target retention and reverse-attention interference evidence & Dedicated AV-TSE \\
Co-occurring-face attention~\cite{pan2025cooccurring} & 2025 & Target and other visible faces & Complementary face-activity evidence & Dedicated AV-TSE \\
Online AV-CrossNet~\cite{yu2025onlineavcrossnet} & 2025 & Synchronized lip motion & Causal frame-level articulatory conditioning & Online AV-TSE \\
AV-ASEN~\cite{avase} & 2025 & Lip motion and past extracted speech & Current articulation and autoregressive acoustic evidence & Online AV-TSE \\
2S-AVTSE~\cite{twostagetse} & 2026 & Lip motion and visual activity & Target activity followed by speech extraction & Online AV-TSE \\
SeLG~\cite{pan2026beyondlips} & 2026 & Lip motion and co-speech gesture & Complementary articulatory and behavioral evidence & Multimodal TSE \\

\midrule
\rowcolor{gray!20}\multicolumn{5}{@{}c}{\rule{0pt}{2.6ex}\small\textbf{Spatial and location-conditioned systems}}\\[0.6ex]
Direction-Aware Speaker Beam~\cite{li2019directionaware} & 2019 & Enrollment utterance and beam bank & Enrollment-guided selection among fixed beams & Spatial TSE \\
NSF~\cite{gu2019neural} & 2019 & Direction and multichannel spatial features & Consistency with a target direction & Spatial TSE \\
ADL-MVDR~\cite{zhang2021adl} & 2021 & Estimated direction & Direction-conditioned masks and beamforming & Spatial TSE \\
Channel-decorrelation TSE~\cite{han2021multichannel} & 2021 & Enrollment and multichannel differences & Speaker adaptation with inter-channel differential information & Spatial TSE \\
3DSF~\cite{gu20213d} & 2021 & Azimuth, elevation, and distance & Three-dimensional target position & Spatial TSE \\
cNSF~\cite{gu2021complex} & 2021 & Direction and interchannel phase & Complex spectral consistency with the target direction & Spatial TSE \\
L-SpEx~\cite{ge2022spex} & 2022 & Estimated direction & Localization-assisted speaker extraction & Spatial TSE \\
LBT$^{\dagger}$~\cite{taherian2022multi} & 2022 & Source location during training & Location-based output assignment & Spatially assigned separation \\
TS-TSE$^{\dagger}$~\cite{xu2022learning} & 2022 & Binaural sector & Membership in a queried spatial sector & Regional target-sound extraction \\
Direction-aware CVAE~\cite{wang2022directioncvae} & 2022 & Target direction & Geometrically constrained target source model & Spatial TSE \\
AN-BF~\cite{gu2022towards} & 2023 & Target azimuth & Direction-conditioned neural beamforming & Spatial TSE \\
NS-Extractor~\cite{lin2023focus} & 2023 & Perceived target distance & Near-field target location and self-enrollment & Spatial TSE \\
BG-TSE~\cite{elminshawi2023beamformer} & 2023 & Enrollment and target-steered beamformer output & Time-varying beamforming guidance & Spatial TSE \\
iCOSPA~\cite{briegleb2023icospa} & 2023 & Target location & Location-informed complex spatial representation & Spatial TSE \\
DoA-assisted TSE~\cite{wang2024study} & 2024 & Estimated or oracle direction & Direction-assisted target assignment & Spatial TSE \\
ReZero$^{\dagger}$~\cite{gu2024rezero} & 2024 & Angular or distance region & Membership in a user-defined spatial region & Regional target-sound extraction \\
BASNet$^{\dagger}$~\cite{yang2024binaural} & 2024 & Binaural angular region & Membership in a target azimuth range & Regional target-sound extraction \\
Binaural selective attention~\cite{meng2024binaural} & 2024 & Enrollment utterance and binaural cues & Speaker identity combined with binaural spatial evidence & Multimodal TSE \\
All-neural directional extraction~\cite{pandey2024directional} & 2024 & Frame-wise direction & Time-varying directional selection & Spatial TSE \\
Distance-based TSE~\cite{shi2025distance} & 2025 & Target distance & Consistency with a queried distance & Spatial TSE \\
M2M-TSE$^{\dagger}$~\cite{choi2025multichannel} & 2025 & Direction and time range & Spatial and temporal target specification & Multichannel target-sound extraction \\
Direction-based BiTSE~\cite{wang2025leveraging} & 2025 & Binaural directional embedding & Target direction in binaural mixtures & Spatial TSE \\
DoA-versus-speaker comparison~\cite{zhang2025doaorspeaker} & 2025 & Direction or enrollment utterance & Controlled comparison of spatial and identity cues & Comparative multimodal TSE \\
DoA-guided extraction$^{\dagger}$~\cite{jing2025doaguided} & 2025 & Direction and beamwidth & Membership in a direction-centered region & Regional target-sound extraction \\
SSDQ~\cite{zhu2025ssdq} & 2025 & Semantic description and spatial region & Joint semantic and location constraints & Multimodal TSE \\
Spectral-or-spatial TSE~\cite{eisenberg2026spectralspatial} & 2026 & Enrollment utterance or direction & Reliability-dependent identity or spatial selection & Multimodal TSE \\
MC-LExt~\cite{ling2026mclext} & 2026 & Onset-prompted enrollment and multichannel audio & Self-enrolled identity with spatial mixture evidence & Spatial TSE \\
\rowcolor{gray!20}\multicolumn{5}{@{}c}{\rule{0pt}{2.6ex}\small\textbf{Textual or semantic systems}}\\[0.6ex]
Heterogeneous TSE~\cite{tzinis2022heterogeneous} & 2022 & Semantic source label & Membership in a requested speech category & Semantic speech extraction \\
Waveformer$^{\dagger}$~\cite{veluri2023real} & 2023 & Class label & Membership in a requested sound class & Target-sound extraction \\
CLAPSep$^{\dagger}$~\cite{ma2024clapsep} & 2024 & Text or audio prompt & Cross-modal semantic similarity & Target-sound extraction \\
DPM-TSE$^{\dagger}$~\cite{hai2024dpm} & 2024 & Sound class label & Class-conditioned diffusion & Target-sound extraction \\
AudioSep$^{\dagger}$~\cite{liu2024separate} & 2025 & Caption or class label & Open-vocabulary sound description & Target-sound extraction \\
SoloAudio$^{\dagger}$~\cite{wang2025soloaudio} & 2025 & Natural-language description & Language-conditioned latent diffusion & Target-sound extraction \\
StyleTSE~\cite{huo2025beyond} & 2025 & Speaking-style description & Para-linguistic target attributes & Text-guided TSE \\
ISRC-TSE~\cite{dai2025inter} & 2025 & Inter-speaker relation & Relative attributes within the current mixture & Text-guided TSE \\
PNTP-TSE~\cite{seki2025language} & 2025 & Para-linguistic or non-linguistic prompt & Prompted target attributes & Text-guided TSE \\
CSE~\cite{kim2025contextual} & 2025 & Dialogue history & Conversational consistency with a candidate speaker & Context-guided TSE \\
pTSE-T~\cite{jiang2024ptse} & 2026 & Unaligned presentation text & Content associated with a presentation speaker & Text-guided TSE \\
LLM-TSE~\cite{hao2025typing} & 2026 & Natural-language speaker description & Speaker attributes and scene descriptions & Text-guided TSE \\
TPEech~\cite{jiang2026tpeech} & 2026 & Dialogue history & Historical text and echo-based contextual evidence & Context-guided TSE \\
\midrule
\rowcolor{gray!20}\multicolumn{5}{@{}c}{\rule{0pt}{2.6ex}\small\textbf{Neural and attention-conditioned systems}}\\[0.6ex]
BISS~\cite{biss2020} & 2020 & EEG or iEEG & Neural response associated with the attended stream & Neural TSE \\
BESD~\cite{hosseini2021} & 2021 & EEG & Time-aligned auditory-attention evidence & Neural TSE \\
U-BESD~\cite{hosseini2022} & 2022 & EEG & Auditory-attention evidence in a U-shaped extractor & Neural TSE \\
BASEN~\cite{zhang2023basen} & 2023 & EEG & Cross-modal neural and speech association & Neural TSE \\
NeuroHeed~\cite{pan2023neuroheed} & 2024 & EEG & Neuronal attractor representing the attended source & Neural TSE \\
NeuroHeed+~\cite{pan2024neuroheed+} & 2024 & EEG & Joint attention detection and extraction & Neural TSE \\
NeuroSpEx~\cite{neurospex} & 2024 & EEG & Cross-attention between neural and speech representations & Neural TSE \\
MSFNet~\cite{MSFNet} & 2024 & EEG and visual observations & Multiscale neural evidence with visual context & Multimodal neural TSE \\ 
M3ANet~\cite{m3anet} & 2025 & EEG & Temporally aligned neural and speech representations & Neural TSE \\
NeuroSpEx+~\cite{neurospex+} & 2025 & EEG & Neural conditioning with attended-envelope supervision & Neural TSE \\
IFENet~\cite{fan2025} & 2025 & EEG & Neural attention encoded with nonlinear feature modeling & Neural TSE \\
MLAD-ETSEN~\cite{wang2025mlad} & 2025 & EEG & Joint extraction and auditory-attention decoding & Neural TSE \\
TIDENet~\cite{lv2026tidenet} & 2026 & EEG & Interpolated and structurally aligned neural evidence & Neural TSE \\
BM-TSE~\cite{han2026brainprint} & 2026 & EEG and learned brainprint & Listener-specific attention and identity characteristics & Personalized neural TSE \\
\end{xltabular}
\setlength{\tabcolsep}{6pt}
\normalsize
\end{landscape}
\twocolumn
\raggedbottom